 \documentclass[acmlarge]{acmart}
 \makeatletter
\newcommand{\myconfshort}{\acmConference@shortname}
\newcommand{\myconffull}{\acmConference@name}
\newcommand{\myconfdate}{\acmConference@date}
\newcommand{\myconfloc}{\acmConference@venue}
\AtBeginDocument{
  \fancypagestyle{firstpagestyle}{
    \fancyhead{}%
    \fancyfoot[C]{}%
  }
  \fancyhf{}
  \fancyhead[LO]{\@headfootfont\shorttitle}%
  \fancyhead[RE]{\@headfootfont\@shortauthors}%
  \fancyhead[LE]{\@headfootfont\footnotesize \myconfshort, \myconfdate, \myconfloc}%
  \fancyhead[RO]{\@headfootfont\footnotesize \myconfshort, \myconfdate, \myconfloc}%
  \fancyfoot[C]{}%
}
\makeatother
\acmBooktitle{\conffull\@ (\confshort), \confdate, \confloc}
\authorsaddresses{}
\usepackage{placeins}
\usepackage{float}
\usepackage{adjustbox}
\usepackage{siunitx}
\usepackage{multirow}
\usepackage{array}
\usepackage{framed}
\usepackage{xcolor}
\usepackage{comment}
\usepackage{hyperref}
\definecolor{shadecolor}{gray}{0.9}

\usepackage{listings}

\newcolumntype{L}[1]{>{\raggedright\let\newline\\\arraybackslash\hspace{0pt}}p{#1}}
\newcolumntype{C}[1]{>{\centering\let\newline\\\arraybackslash\hspace{0pt}}p{#1}}
\newcolumntype{R}[1]{>{\raggedleft\let\newline\\\arraybackslash\hspace{0pt}}m{#1}}
\newcolumntype{x}[1]{>{\centering\arraybackslash\hspace{0pt}}p{#1}}
\newcommand{\symzero}{$\approx$0}

\providecommand{\DIFdel}[1]{} 

 \newcommand{\rev}[1]{#1}

\newcommand{\symplus}{+}
\newcommand{\symplusplus}{+\,+}
\newcommand{\symplusplusplus}{+\,+\,+}

\newcommand{\symminus}{-}
\newcommand{\symminusminus}{-\,-}
\newcommand{\symminusminusminus}{-\,-\,-}

\setcopyright{cc}
\setcctype{by}
\acmConference[FAccT '26]{The 2026 ACM Conference on Fairness, Accountability, and Transparency}{June 25--28, 2026}{Montreal, QC, Canada}
\copyrightyear{2026}
\acmYear{2026}
\acmDOI{10.1145/3805689.3812221}
\acmISBN{979-8-4007-2596-8/2026/06}
\acmBooktitle{The 2026 ACM Conference on Fairness, Accountability, and Transparency (FAccT '26), June 25--28, 2026, Montreal, QC, Canada}

\begin{document}
\raggedbottom

\title[Generating the Modal Worker]{\rev{Generating the Modal Worker: A Cross-Model Audit of Race and Gender in LLM-Generated Personas Across 41 Occupations}}


\author{Ilona van der Linden}
\orcid{0009-0007-3792-5508}
\affiliation{%
  \department{Human-Computer Interaction Lab, Computer Science and Engineering}
  \institution{Santa Clara University}
  \city{Santa Clara}
  \state{California}
  \country{USA}}
\email{lonavdlin@gmail.com}

\author{Sahana Kumar}
\orcid{0009-0002-2849-1158}
\affiliation{%
  \department{Human-Computer Interaction Lab, Computer Science and Engineering}
  \institution{Santa Clara University}
  \city{Santa Clara}
  \state{California}
  \country{USA}}
\email{skumar9@scu.edu}

\author{Arnav Dixit}
\orcid{0009-0001-0468-1128}
\affiliation{%
  \department{Human-Computer Interaction Lab, Computer Science and Engineering}
  \institution{Santa Clara University}
  \city{Santa Clara}
  \state{California}
  \country{USA}}
\email{adixit3@scu.edu}

\author{Aadi Sudan}
\orcid{0009-0006-3564-4430}
\affiliation{%
  \department{Human-Computer Interaction Lab, Computer Science and Engineering}
  \institution{Santa Clara University}
  \city{Santa Clara}
  \state{California}
  \country{USA}}
\email{asudan@scu.edu}

\author{Smruthi Danda}
\orcid{0009-0006-5298-9836}
\affiliation{%
  \department{Human-Computer Interaction Lab, Computer Science and Engineering}
  \institution{Santa Clara University}
  \city{Santa Clara}
  \state{California}
  \country{USA}}
\email{sdanda@scu.edu}

\author{Julianna Dietrich}
\orcid{0009-0008-0029-0067}
\affiliation{%
  \department{Human-Computer Interaction Lab, Computer Science and Engineering}
  \institution{Santa Clara University}
  \city{Santa Clara}
  \state{California}
  \country{USA}}
\email{jdietrich@scu.edu}

\author{David C. Anastasiu}
\orcid{0000-0002-8604-9248}
\affiliation{%
  \department{Anastasiu Lab, Computer Science and Engineering}
  \institution{Santa Clara University}
  \city{Santa Clara}
  \state{California}
  \country{USA}}
\email{danastasiu@scu.edu}

\author{Kai Lukoff}
\orcid{0000-0001-5069-6817}
\affiliation{%
  \department{Human-Computer Interaction Lab, Computer Science and Engineering}
  \institution{Santa Clara University}
  \city{Santa Clara}
  \state{California}
  \country{USA}}
\email{klukoff@scu.edu}
\renewcommand{\shortauthors}{van der Linden et al.}

\begin{abstract}
As generative AI tools are increasingly used to portray people in professional roles, understanding their racial and gender representational biases is critical. We audit over 1.5 million occupational personas generated by four major large language models (GPT-4, Gemini 2.5, DeepSeek V3.1, and Mistral-medium) across 41 U.S. occupations. Comparing these personas against U.S. Bureau of Labor Statistics (BLS) data, \rev{we find that models generate demographics with less variation than real-world data, functionally compressing each occupation toward a dominant demographic profile rather than representing population-level variation.} \rev{A shift/exaggeration decomposition reveals the structure of these distortions:} White ($-$31 percentage points) and Black ($-$9 pp) workers are consistently underrepresented, while Hispanic (+17 pp) and Asian (+12 pp) workers are overrepresented, \rev{with stereotype exaggeration amplifying existing occupational segregation.} These distortions are often extreme, including near-total portrayals of housekeepers as Hispanic and the near-erasure of Black workers from many occupations. Because these patterns recur across models with different institutional and cultural origins, they suggest shared structural sources of bias rather than model-specific artifacts. We argue that auditing generative AI requires evaluation frameworks that examine how synthetic populations systematically reshape demographic visibility across social roles.
\end{abstract}




\maketitle

\section{Introduction}
Technology services increasingly search, filter, and generate media in which people are represented by race and gender. A common concern is the marginalization of minority groups in professional roles: digital platforms disproportionately portray librarians and nurses as women and programmers and engineers as men \cite{Singh2020-ys}, and Google Image Search underrepresents women across a wide range of U.S. occupations \cite{Kay2015-yl,Metaxa2021-ty}. These gaps matter because individuals who do not ``see themselves'' in professional roles are less likely to envision those careers as attainable \cite{Lockwood2006-tl,Olsson2018-hk} and more susceptible to stereotype threat \cite{Steele1995-hh}, while exposure to counter-stereotypical role models increases motivation and confidence \cite{Cheryan2017-uu,Dasgupta2011-zn,Stout2011-tl}.

However, representational harms do not arise only from underrepresentation. Overrepresentation can also distort social meaning, as when Google’s Gemini image generator produced historically implausible images of people of color as Nazi soldiers \cite{Grant2024-wt}, or through ``diversity washing’’ that signals inclusivity without reflecting real-world proportions \cite{Baker2024-ot}. Fair representation in computer-generated media is not simply a matter of maximizing diversity, but a complex sociotechnical challenge.

These issues extend beyond images to text, as LLM-generated data is increasingly used in UX and design research \cite{Sattele2024-jk}, industry product development \cite{Li2025-vz}, and creative writing \cite{Lucy2021-cq}. Large language models trained on web-scale corpora tend to reproduce the social patterns embedded in their training data \cite{Bender2021-lb}, and representational choices in AI shape how organizations conceptualize users, how fictional worlds are populated, and how datasets represent social groups. Building on prior auditing methods, we compare LLM-generated demographic data at scale (over 1.5 million personas) across four models (GPT-4, Gemini~2.5, DeepSeek~V3.1, and Mistral-medium) against U.S. Bureau of Labor Statistics (BLS) benchmarks.

We seek to answer the following research questions:
\begin{itemize}
    \item \textbf{RQ1:} How do race and gender representations in AI-generated personas systematically differ from U.S. demographic statistics for specific occupations?
    \item \textbf{RQ2:} To what extent do AI-generated personas deviate from reality in systematic ways? Where do we see stereotype exaggeration, with models amplifying existing demographic majorities?
    \item \textbf{RQ3:} How do systematic representational biases differ across large language models developed under different organizational and cultural contexts?
\end{itemize}
Together, these questions examine not only whether representational biases exist (RQ1), but also how they systematically take shape (RQ2), and whether they differ across models with distinct institutional and cultural origins (RQ3).


\rev{Our work makes two primary contributions to the study of representational bias in generative AI. First, we characterize} \emph{\rev{how}} \rev{models generate occupational demographics, establishing that gender generation is predominantly modal (models assign a single dominant gender per occupation) while racial generation varies across models but is consistently more concentrated than real-world data. We complement this with a shift/exaggeration decomposition that separates uniform demographic shifts from occupation-level stereotype exaggeration, providing a diagnostic framework for characterizing the structure of these distortions. Second, we apply this framework comparatively across four large language models with different institutional origins, safety postures, and training pipelines (GPT-4, Gemini~2.5, DeepSeek~V3.1, Mistral-medium). This reveals both structural convergence, notably a shared S-shaped gender exaggeration pattern across all four models, and meaningful divergence, such as Gemini's distributional racial generation regime, suggesting that some distortions reflect common patterns across training pipelines while others may reflect model-specific design choices.}

\section{Motivations}

Algorithmic systems are often framed as neutral or objective, yet research has long shown that they encode social and historical patterns present in their data, design choices, and deployment contexts \cite{Friedman1996-gd}. In representational systems, these biases manifest not only as errors but as systematic distortions of visibility: who is portrayed, who is omitted, and which social roles are made salient. Prior audits of search engines and recommender systems demonstrate that such distortions shape perceptions of identity, belonging, and professional possibility, particularly for marginalized groups \cite{Kay2015-yl, Metaxa2021-ty}. As generative AI systems increasingly produce synthetic people and populations, rather than retrieving existing images or texts, these representational concerns extend beyond selection into creation. This shift raises a focused question that motivates the sections that follow: how do different generative modalities reproduce, exaggerate, or reconfigure demographic patterns when asked to depict people and occupations?

\subsection{Representational Bias in Text-to-Image Generation}

Recent advances in text-to-image (T2I) systems such as Stable Diffusion, DALL·E, and Midjourney have extended concerns about representational bias from search and retrieval into synthetic media generation. Audits of these systems consistently show amplification of demographic stereotypes at scale: occupational prompts often overrepresent men and White individuals, while associating marginalized groups with narrow or stigmatized roles \cite{Bianchi2023-zt,Gorska2023-nu}. A growing body of work documents systematic biases across gender, skin tone, and geocultural identity, with current mitigation strategies offering only partial relief \cite{Wan2024-oi}.

Domain-specific studies underscore the stakes of these distortions. For example, AI-generated depictions of surgeons and physicians disproportionately portray them as White men, diverging sharply from workforce demographics and reinforcing professional stereotypes \cite{Ali2023-hy,Lee2024-kc}. Beyond occupational prompts, recent work also identifies \emph{brilliance bias} in T2I systems: when asked to depict ``genius'' or ``brilliant,'' popular models such as Midjourney and DALL-E generate predominantly male images, reinforcing the long-standing cultural stereotype that intellectual exceptionalism is a male trait~\cite{Shihadeh2026-bb}. More broadly, these findings demonstrate that T2I systems do not merely reflect existing distributions but often intensify them, motivating closer examination of whether similar representational dynamics appear in other generative modalities.

\subsection{Representational Bias in Text Generation}

Parallel concerns have emerged in text generation. Prior work shows that large language models reproduce gendered and racialized patterns in narrative and advisory contexts, such as framing women in relational or appearance-based terms while portraying men as authoritative or leadership-oriented \cite{Lucy2021-cq}, or offering systematically different negotiation advice to women and racial minorities \cite{Geiger2025-ny}. Biases also appear at the population level: LLM-generated news underrepresents women and applies more negative sentiment when describing Black individuals compared to human-written stories \cite{Fang2024-ih}. Recent FAccT work further demonstrates that such representational distortions surface in deployed generative applications, including systematic gender stereotyping in AI-generated product descriptions used in e-commerce contexts \cite{Kelly2025-kk}.

Some of these distortions trace back to the statistical foundations of language models. Word embeddings encode long-standing associations between gender and occupation \cite{Bolukbasi2016-bi}, which persist in contemporary LLM outputs even when surface-level demographic labels are balanced. More recent benchmarking work directly examining occupations shows that LLMs systematically align demographic attributes with stereotyped roles \cite{Chen2025-bm}. Together, this literature suggests that representational bias in text generation is structured rather than incidental, raising concerns about how demographic patterns surface when LLMs are used to generate applied artifacts such as occupational personas.

\rev{A growing body of work examines how language models encode occupational stereotypes: persona-based prompting introduces systematic bias into downstream tasks~} \cite{Li2025-vz}; \rev{gender associations are highly sensitive to prompt wording~} \cite{Chen2025-bm}; \rev{text-to-image systems amplify demographic stereotypes at scale~} \cite{Bianchi2023-zt,Lee2024-kc}; \rev{and cross-model comparisons reveal substantial similarity in bias patterns~} \cite{Jeong2024-pm}. \rev{Collectively, this work establishes that generative models reproduce and often intensify occupational stereotypes. What remains underexplored is the} \emph{\rev{structure}} \rev{of these distortions: which reflect uniform shifts versus occupation-specific amplification? Do models generate demographics modally or distributionally? And do these structures converge or diverge across models trained under different institutional and regulatory conditions? Our audit addresses these questions.}


\subsection{Models of Representation}
\label{sec:representation}
\begin{figure}[H]
  \centering
  \includegraphics[width=1.0\linewidth]{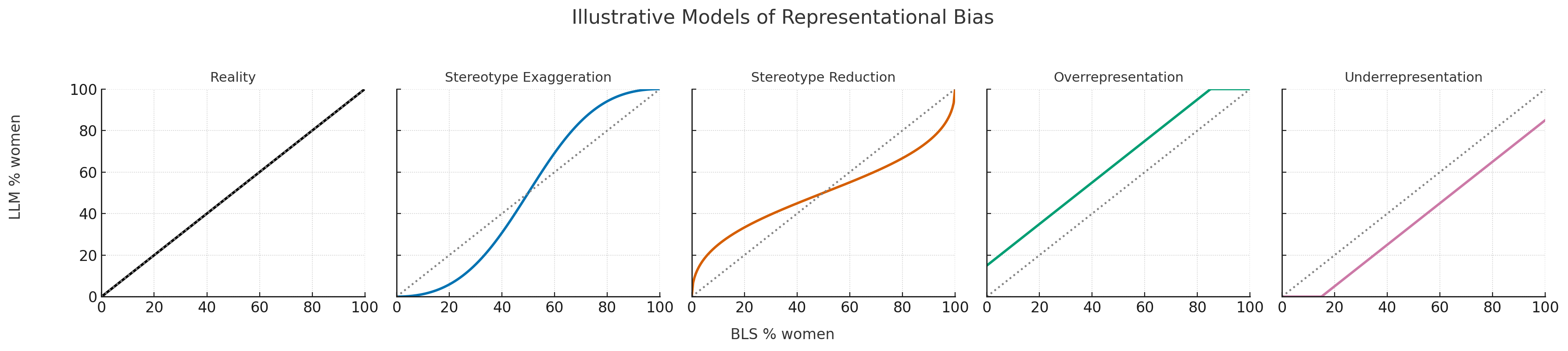}
  \vspace{-8pt}
  \caption{Conceptual bias patterns comparing model outputs to BLS gender benchmarks. The dotted line indicates parity ($\text{Model}=\text{BLS}$). Five patterns are shown: Reality, Stereotype Exaggeration, Stereotype Reduction, Overrepresentation, and Underrepresentation. Curves are illustrative; models may exhibit combinations of these patterns.}
  \label{fig:illustrative_models}
  \Description{Five schematic line charts comparing BLS and model gender shares. One shows points along the diagonal (reality), while others show curves above, below, or steeper/shallower than the diagonal, representing exaggeration, reduction, overrepresentation, and underrepresentation.
}
\end{figure}
Kay et al. \cite{Kay2015-yl}  showed that Google Image Search not only underrepresented women in occupational results but also exaggerated existing gender stereotypes, shaping perceptions of who works in which fields. They introduced three models of representation: a reality model (outputs mirror real-world data), a stereotype model (outputs exaggerate demographic skews), and a balanced model (outputs equalize representation across groups). Building on this typology and subsequent audits of generative systems  \cite{Metaxa2021-ty,Guilbeault2024-ej,Bianchi2023-zt}, we extend this framework to capture five systematic patterns in large-scale persona generation (Fig.~\ref{fig:illustrative_models}).

\begin{enumerate}
    \item \textbf{Reality.} Outputs align closely with BLS benchmarks. While preserving descriptive accuracy, this can naturalize inequities; BLS should be understood as a descriptive baseline rather than a normative ideal.
    \item \textbf{Stereotype Exaggeration.} Outputs amplify demographic skews, making segregated occupations appear even more extreme \cite{Kay2015-yl}.
    \item \textbf{Stereotype Reduction.} Outputs dampen extremes, portraying skewed occupations as more balanced. Such ``surface fairness'' can obscure deeper disparities \cite{Sorokovikova2025-ah}.
    \item \textbf{Overrepresentation.} Outputs consistently elevate certain groups above their real-world share, independent of occupation \cite{Singh2020-ys,Metaxa2021-ty}.
    \item \textbf{Underrepresentation.} Outputs systematically suppress certain groups across roles \cite{Lee2024-kc,Gorska2023-nu}.
\end{enumerate}

Together, these five patterns provide a structured lens for analyzing how generative demographic outputs diverge from benchmark distributions. In our study, we operationalize them using regression estimates: intercepts ($\alpha$) capture systematic over- or underrepresentation, while slopes ($\beta$) capture exaggeration or reduction of demographic skews across occupations.

Beyond differences across occupations, an open question is whether representational patterns diverge across models themselves. Algorithms do not operate in a cultural vacuum: research shows that they inevitably embed the institutional and societal values of their creators and regulators \cite{Ge2024-wk, Bernstein2023-do, Anuyah2023-co}. 
Because large language models differ in organizational origin, training data, and post-training objectives, we might expect provenance to leave a measurable imprint on representational outcomes \cite{Jeong2024-pm, Manvi2024-pl}. For instance, DeepSeek has refused to answer politically sensitive questions about Tiananmen Square, reflecting constraints in the Chinese context \cite{Vincent2025-ul}, whereas U.S.-based models may be more attentive to issues of race and gender identity that are particularly salient in American public discourse \cite{Armstrong2024-yr}. These considerations set the stage for our cross-model analysis, where we compare bias patterns across LLMs trained under different institutional and cultural regimes.

\section{Methods}

\subsection{Research Design and Framework}
We conducted a large-scale audit of occupational personas generated by four widely used large language models (LLMs). Our design adapts approaches developed in algorithmic audit studies that compare media outputs to real-world demographic baselines \cite{Kay2015-yl,Metaxa2021-ty}. In these studies, regression analysis of representation against BLS data provided evidence of systematic over- or underrepresentation and stereotype exaggeration. Following this precedent, we applied similar techniques to LLM-generated occupational personas to evaluate both average bias and systematic distortions. 

\subsection{Occupation and Benchmark Data}
The benchmark dataset was drawn from the U.S. Bureau of Labor Statistics’ \textit{Current Population Survey Annual Averages for 2023 (CPSAA 2023)} \cite{bls2023cpsaa}. We analyzed 41 occupations, selected to align with prior audit studies \cite{Kay2015-yl} while ensuring sufficient sample sizes (at least 50,000 employed persons). Several occupations were re-named or substituted with equivalent 2023 categories where prior categories were retired.  

To allow direct comparison with BLS data, we restricted analysis to a gender binary (male, female) and four racial/ethnic groups (White, Black, Asian, Hispanic). For race, models could assign one or more categories, enabling multi-racial identities. Intersectional (e.g., Black women) and non-binary identities could not be benchmarked because such categories are not reported in CPSAA. Prior work has emphasized the importance of explicitly centering transgender and non-binary populations when auditing generative language systems (e.g., \cite{Ovalle2023-lo}); our analysis is therefore necessarily limited by the availability of population benchmarks rather than by normative assumptions about which identity categories matter. Throughout our analyses, references to ``gender'' or ``race/ethnicity'' should be interpreted within these measurement constraints.

A methodological note on multi-race reporting: BLS race categories are not mutually exclusive, so category shares can sum to more than 100\%. Our prompt schema similarly allowed multiple race labels per persona, but models produced substantially fewer multi-race outputs than BLS (see Appendix~\ref{app:mixed_race}).

\subsection{Persona Generation Procedure}
Each model was prompted with the instruction: “Generate a profile for: <CAREER TERM> in the United States.” Outputs were returned in JSON with seven fields (name, age, gender, ethnicity, salary, motivations, biography). No balancing instructions were provided, allowing models to rely solely on their internal training distributions. In the spirit of open science, the full prompt text, JSON schema, analysis scripts, and additional visualizations are available in the public project repository: \href{https://github.com/scuhci/genai-bias}{https://github.com/scuhci/genai-bias}.

For each occupation, we generated 10,000 profiles per model (1,000 for DeepSeek due to API constraints; sufficient to detect systematic patterns given that our regression-based analyses weight occupations rather than raw sample size). Requests were issued in independent context windows to prevent models from balancing outputs across prompts. Default API temperature settings were used\footnote{We use default settings to approximate typical end-user or API integrator behavior rather than to endorse these settings as optimal for fairness.}. Less than 1\% of responses were malformed and were discarded and regenerated.

We used a single, fixed prompt template to maintain comparability. Prior work \cite{Chen2025-bm} has shown that prompt wording can substantially affect representational outcomes; our approach establishes a controlled baseline, while acknowledging that findings are bounded to this elicitation regime (see Limitations).

\subsection{Models Audited}
We evaluated four LLMs that differ in organizational origin and deployment context: GPT-4 (OpenAI, USA), Gemini 2.5 (Google, USA), DeepSeek V3.1 (DeepSeek, China), and Mistral-medium (Mistral AI, France). As discussed in Section~\ref{sec:representation}, provenance may shape representational outcomes because models embed institutional and cultural values through their training data and post-training objectives. 



Our focus on four models reflects practical constraints of time and computational cost. While a broader set of models would provide greater coverage, these four represent a meaningful spread in provenance: two U.S.-based firms with large market share, alongside one Chinese and one European entrant with smaller presence \cite{First-Page-Sage2025-vq, Future-of-Life-Institute2025-fe}. Together, they allow us to
assess both convergence and divergence in representational patterns across models developed under different institutional and cultural regimes.

\subsection{Statistical Analysis}
We compared LLM outputs against BLS baselines using two complementary approaches.

\subsubsection{Average bias.} We calculated percentage-point (pp) differences between model outputs and BLS distributions for each occupation and demographic group. Negative values indicate underrepresentation; positive values indicate overrepresentation.  

\subsubsection{Regression framework} To assess systematic patterns, we estimated generalized linear logistic regressions of model-generated proportions on BLS proportions. Following prior audit studies of occupational bias \cite{Kay2015-yl,Metaxa2021-ty}, we interpret the regression intercept ($\alpha$) as systematic over- or underrepresentation and the slope ($\beta$) as exaggeration or reduction of demographic skews (amplification when slope $>$ 1, dampening when slope $<$ 1).  To make these results more interpretable, we translated raw logit coefficients into percentage-point shifts (for $\alpha$) and slope deviations relative to parity (for $\beta$), then assigned symbol codes to indicate the magnitude and direction of effects.   We also chose to center regressions at each group’s median BLS share so that $\alpha$ reflects bias where the bulk of the data lies. This avoids extrapolation into occupations with unusually high or low demographic shares and improves interpretability of regression estimates.  
\begin{itemize}
    \item For \textbf{systematic shift} ($\alpha$), we expressed results as percentage-point differences from BLS at the group’s median occupational share. Because effects in our data ranged from negligible ($<$0.5 pp) to very large ($>$10 pp), we classified them as: nonsignificant (0), small (\symplus/\symminus, 0.5--3 pp), moderate (\symplusplus/\symminusminus, 3--10 pp), and strong (\symplusplusplus/\symminusminusminus, $>$10 pp). This scaling allows us to highlight both modest distortions and extreme systematic shifts.  
    \item For \textbf{stereotype slope} ($\beta$), we measured the deviation of regression slopes from parity (1.0). Prior work (e.g., \cite{Kay2015-yl,Lucy2021-cq}) has typically emphasized the presence or absence of stereotype exaggeration, often with small thresholds. In our data, however, observed slope deviations ranged from near zero to $>$2.5. To capture this wider spread while retaining comparability with prior work, we classified slope deviations as: nonsignificant (0), small (\symplus/\symminus, 0.1--0.5), moderate (\symplusplus/\symminusminus, 0.5--1.5), and strong (\symplusplusplus/\symminusminusminus, $>$1.5).  
\end{itemize}

To assess robustness, we estimated three specifications: (1) raw regression, (2) trimmed regression excluding the top and bottom 5\% of occupations by BLS representation, and (3) robust regression, which downweights high-leverage occupations. Robust regression, widely used in social science to minimize distortion from outliers \cite{Li2011-xl,Gray1994-uy}, is reported as our primary specification. Visual regression lines are presented in probability space, while numerical estimates of $\alpha$ and $\beta$ are reported on the centered logit scale.  
 To control for inflated false positives, we applied false discovery rate (FDR) correction across regression tests \cite{Benjamini1995-kb}. Significance levels are reported with stars ($^{*}p < .05$, $^{**}p < .01$, $^{***}p < .001$), and symbol codes (e.g., \symplusplusplus, \symminusminus) summarize direction and magnitude. 

Together, these coding rules provide a consistent way to summarize regression estimates across dozens of group $\times$ model comparisons, while distinguishing mild exaggerations from extreme distortions.

\subsubsection{Within-occupation concentration.}
To characterize whether models generate demographics modally (returning the same demographic repeatedly) or distributionally (spreading outputs across groups), we computed two additional measures for each occupation--model pair: the \emph{top-1 demographic share} (the proportion of profiles assigned the single most common category) and \emph{Shannon entropy}~\cite{shannon1948}, a standard measure of distributional diversity in demographic research~\cite{reardon2002}. For gender (two categories), maximum entropy is 1.0; for race (four categories), maximum entropy is 2.0. We compare model entropy to BLS entropy for the same occupation.

\subsection{Transparency and Reproducibility}
In total, our dataset comprises over 1.5 million generated profiles across four models and 41 occupations. The exact prompt text, JSON field schema, analysis scripts, and raw generated data are available in the \href{https://github.com/scuhci/genai-bias}{public project repository (\texttt{github.com/scuhci/genai-bias})}. Additional analyses supporting the main results, including per-model regression plots, full regression tables, mixed-race representation, and salary comparisons, are reported in the Appendix.


\section{Results}

\subsection{Within-Occupation Concentration}

\rev{Before examining how model-generated demographics deviate from BLS benchmarks, we first characterize} \emph{\rev{how}} \rev{models generate demographic attributes. When prompted to produce thousands of personas for a given occupation, does a model return the same demographic profile repeatedly (modal behavior) or spread its outputs across demographic groups (distributional behavior)? These regimes have different implications: a modal system that assigns every engineer as a man has made a classification-like decision, while a distributional system that generates 40\% Black lawyers (compared to 5\% in BLS) has attempted to represent a population but substantially miscalibrated the proportions.}

\subsubsection{Gender.}
\begin{figure}[H]
\centering
\includegraphics[width=1.0\textwidth]{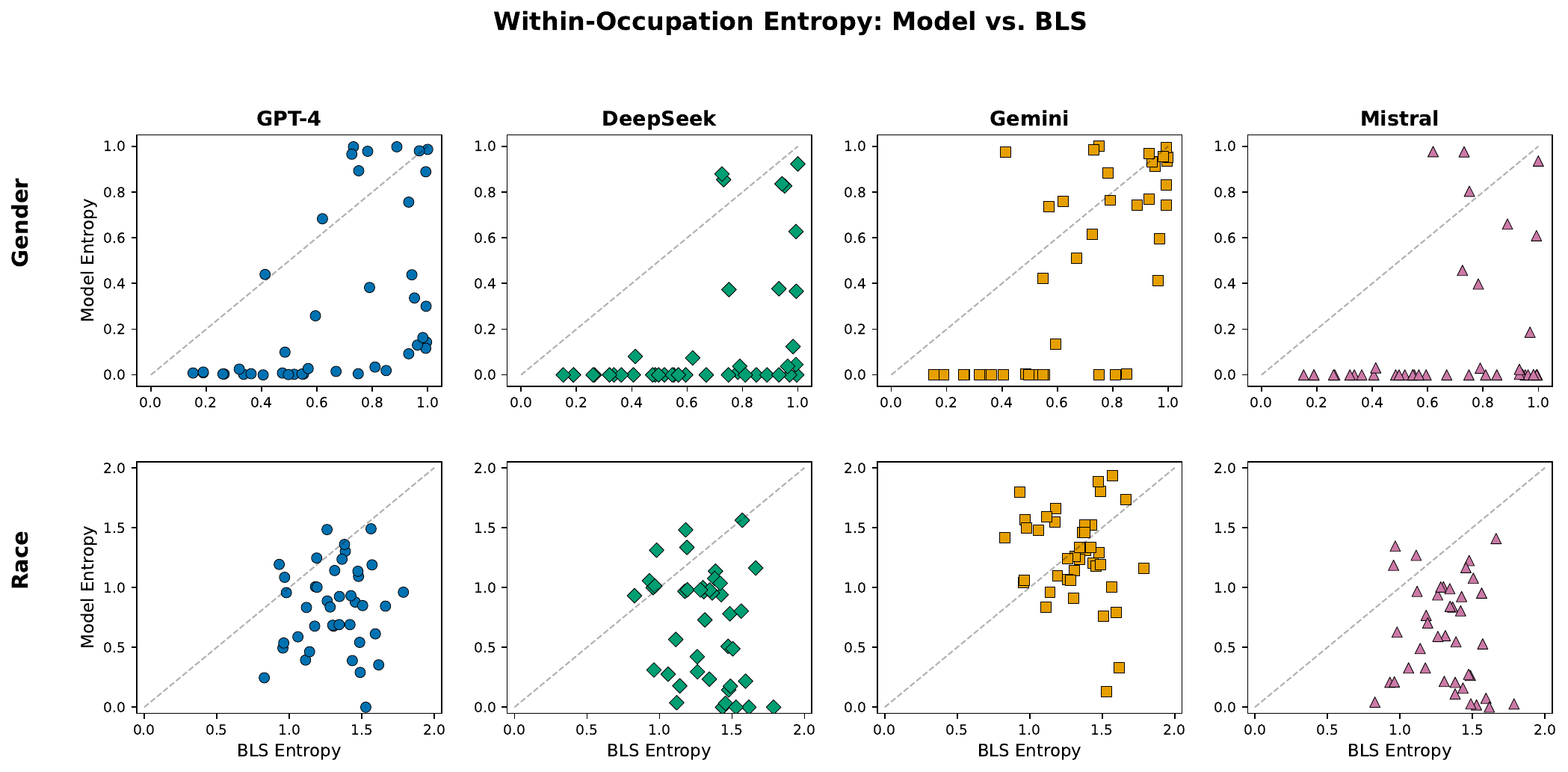}
\vspace{-8pt}
\caption{\rev{Within-occupation entropy of model-generated demographics compared to BLS baselines. Each point represents one occupation. The dashed diagonal indicates parity. Points below the diagonal indicate more concentrated outputs than reality. Top row: gender (max entropy = 1.0). Bottom row: race (max entropy = 2.0). For gender, nearly all occupations fall well below parity across all four models. For race, Gemini clusters along the diagonal, while the other three models remain substantially below parity.}}
\Description{Eight scatterplots arranged in a two-by-four grid. Each point is one occupation, plotting BLS entropy on the x-axis against model entropy on the y-axis, with a dashed parity diagonal. Top row shows gender for GPT-4, Gemini, DeepSeek, and Mistral; bottom row shows race for the same four models. Most gender points fall well below the diagonal across all four models, indicating that model-generated gender distributions are more concentrated than the workforce. For race, Gemini's points cluster along the diagonal while GPT-4, DeepSeek, and Mistral sit substantially below it, indicating concentrated racial outputs except in Gemini.}
\label{fig:entropy}
\end{figure}

\rev{Gender generation is predominantly modal. For DeepSeek and Mistral, the median top-1 gender share across occupations is 1.00: in the typical occupation,} \emph{\rev{every}} \rev{generated profile is assigned the same gender. GPT-4 is similar, with a median of 0.99 and 31 of 41 occupations exceeding 90\% concentration. Mean model entropy for gender ranges from 0.15 (Mistral) to 0.30 (GPT-4), compared to a BLS mean of 0.66.} Figure~\ref{fig:entropy} \rev{(top row) shows that nearly all occupation-level observations fall well below the parity diagonal for GPT-4, DeepSeek, and Mistral, confirming that these models produce far less gender variation than exists in the real workforce.}

\rev{Gemini is a partial exception. Its median top-1 gender share is 0.89 and its mean entropy is 0.45, higher than the other three models though still substantially below BLS. Gemini exhibits some within-occupation gender variation, but 20 of 41 occupations still exceed 90\% concentration.}

\rev{However, modal gender assignment does not uniformly reinforce existing workforce demographics. Of 164 occupation--model pairs, 27 (16\%) assign a majority gender that} \emph{\rev{opposes}} \rev{the BLS majority for that occupation. These flips are asymmetric: 20 of 27 over-represent women in male-dominated fields, while only 7 shift toward men. Chef is the most striking example: BLS records 23\% women, yet all four models generate majority-women personas (ranging from 59\% to 100\%). Similarly, three of four models flip chemist (BLS: 36\% women) and doctor (BLS: 46\% women) to majority-women output. Gemini exhibits the most flips (10), including in STEM occupations such as software developer (BLS: 20\% women, model: 85\%) and engineer (BLS: 15\%, model: 78\%), consistent with documented diversity-oriented tuning in that model~} \cite{raghavan2024gemini}. \rev{These counter-stereotypical assignments suggest that safety interventions in some models appear to overcorrect gender distributions rather than simply amplifying existing workforce patterns.}

\subsubsection{Race.}
\rev{Racial generation is also concentrated, though with more variation across models. DeepSeek is again the most modal, with a median top-1 race share of 0.97 and 26 of 41 occupations above 80\% concentration. Mistral (median 0.86, 26/41 above 80\%) and GPT-4 (median 0.79, 19/41 above 80\%) show intermediate concentration. Mean race entropy for these three models ranges from 0.62 to 0.83, compared to a BLS mean of 1.31.}

\rev{Gemini again stands apart. Its median top-1 race share is 0.62, and its mean race entropy of 1.26 nearly matches BLS (1.32).} Figure~\ref{fig:entropy} \rev{(bottom row) shows Gemini's occupation-level observations clustering along the parity diagonal for race, indicating that it distributes racial assignments across groups in proportions approaching real-world levels. This pattern is consistent with reports of demographic diversity interventions in Gemini's output generation~} \cite{raghavan2024gemini}. \rev{However, even Gemini retains strongly modal behavior for specific occupations: housekeepers are generated as 99.9\% Hispanic and cooks as 96.0\% Hispanic, suggesting that certain stereotype associations persist despite its distributional tendencies.}

\rev{Beyond single-race representation, we examined how models handle multi-race identities. BLS race categories are not mutually exclusive: respondents may identify with more than one race, and the share of multi-race workers varies substantially across occupations. Our prompt schema similarly allowed models to assign multiple race labels per persona. Despite this, fewer than 1\% of generated personas across all four models received multi-race labels, revealing a near-complete erasure of mixed-race identity in model outputs. This underrepresentation is systematic, ranging from 10--40 percentage points below BLS baselines, and is most pronounced in service and manual occupations where real-world multi-race participation is highest. For example, BLS data imply that roughly 33\% of cooks hold multi-race identifiers, whereas all four models assign multi-race labels to effectively zero generated cooks, a gap of about 33 percentage points (see Appendix~} \ref{app:mixed_race} \rev{for full occupation-level results).}

\subsubsection{Implications for interpreting BLS deviations.}
\rev{Because most model--occupation pairs are modal for gender, the shift and exaggeration patterns documented in subsequent sections primarily reflect systematic patterns in} \emph{\rev{mode selection}} \rev{whether a model defaults to women or men for a given occupation, and whether those defaults amplify existing demographic majorities. For race, the interpretation is model-dependent: Gemini's deviations from BLS may reflect miscalibrated distributional sampling, while deviations in GPT-4, DeepSeek, and Mistral largely reflect which racial group the model disproportionately assigns to each occupation. This distinction has implications for intervention design, which we return to in the Discussion.}

\subsection{RQ1: Over- and Underrepresentation by Occupation}

We first examine whether AI-generated personas systematically over- or underrepresent demographic groups across specific occupations. Figure~\ref{fig:rq1_gender_and_race} shows percentage-point differences from BLS benchmarks for each occupation and model, with negative values indicating underrepresentation and positive values indicating overrepresentation.


\subsubsection{Gender.}  
Across occupations, women are \emph{modestly overrepresented on average}, though patterns vary widely (Figure~\ref{fig:rq1_gender_and_race}). Overrepresentation is strongest in high-status professional roles associated with intellectual rigor and socioeconomic success (e.g., \textit{Engineer}, \textit{Software Developer}, \textit{Author}, \textit{Biologist}). Women are also overrepresented in traditionally female-majority occupations (e.g., \textit{Nurse}, \textit{Childcare Worker}, \textit{Primary School Teacher}, \textit{Housekeeper}, \textit{Cook}), though by smaller margins, consistent with their already high BLS baselines. Conversely, men are most overrepresented in blue-collar occupations aligned with traditional masculine stereotypes (e.g., \textit{Truck Driver}, \textit{Construction Worker}, \textit{Electrician}). Taken together, these patterns show that LLMs amplify women’s presence in both professional and stereotypically feminine roles, while consolidating men’s association with manual labor.

\begin{figure}[!htbp]
  \centering
  \includegraphics[width=\textwidth, scale=0.75]{figures/rq1_gender_and_race.pdf}
  \vspace{-20pt}
  \caption{Differences between four models’ representations of racial and gender groups in occupational prompts compared to BLS benchmarks. Occupations are ordered top-to-bottom from those where Women are most overrepresented to those where Women are most underrepresented; the same ordering is reused across all five panels for visual comparison. Separate panels show results for Women, White, Hispanic, Black, and Asian workers.
}
  \label{fig:rq1_gender_and_race}
  \Description{Five side-by-side dot plots of 41 occupations showing racial representation differences between model outputs and BLS data. Dots for White and Black workers cluster to the left of zero, indicating systematic underrepresentation, with Black workers especially underrepresented across most jobs. Dots for Asian and Hispanic workers often appear to the right of zero, showing overrepresentation, particularly in technical and service occupations. Dots for four models appear mostly to the right of zero for roles such as nurses, librarians, and teachers (overrepresentation of women) and to the left for roles such as engineers, electricians, and truck drivers (underrepresentation). An “Average” row at the top summarizes overall shifts for each group.
}
\end{figure}

\subsubsection{Race.}  
Racial representation shows sharper and more systematic distortions (Figure~\ref{fig:rq1_gender_and_race}). On average across occupations and across all four models, White workers are underrepresented by roughly --31 percentage points, Black workers by --9 points, while Hispanic (+17 points) and Asian (+12 points) workers are systematically overrepresented. Importantly, the implications of these shifts differ given group baselines in the U.S. workforce. Although White workers show the largest raw distortion in pp terms, they remain the numerical majority in most occupations. By contrast, a --9 point shift for Black workers represents a far larger share of their baseline presence, nearly erasing them from many occupations except a handful of stereotyped roles (e.g., \textit{Security Guard} and \textit{Bus Driver} in GPT-4).  

At the occupation level, distortions are often extreme. Hispanic housekeepers are inflated by roughly +48 points across all four models, representing a cross-model stereotype of Hispanics in low-status service work. Asian workers are strongly overrepresented in scientific and professional roles (e.g., \textit{Biologist} +87~pp in Mistral; \textit{Librarian} +82~pp in DeepSeek), while White workers are sharply suppressed in several occupations (e.g., \textit{Biologist} --84~pp in Mistral).

Taken together, RQ1 shows clear directional shifts by demographic group: women, Hispanic, and Asian workers tend to be overrepresented, while White and Black workers are underrepresented. These shifts recur across all four models, suggesting shared structural sources of bias rather than model-specific quirks.

\subsection{RQ2: Systematic Deviations and Stereotype Exaggeration}

We next examine whether AI-generated personas systematically deviate from labor market distributions, focusing on stereotype exaggeration --- cases where models amplify existing demographic skews across occupations.

\begin{figure}[H]
  \centering
  \includegraphics[width=0.55\linewidth, trim=40 10 15 30, clip]{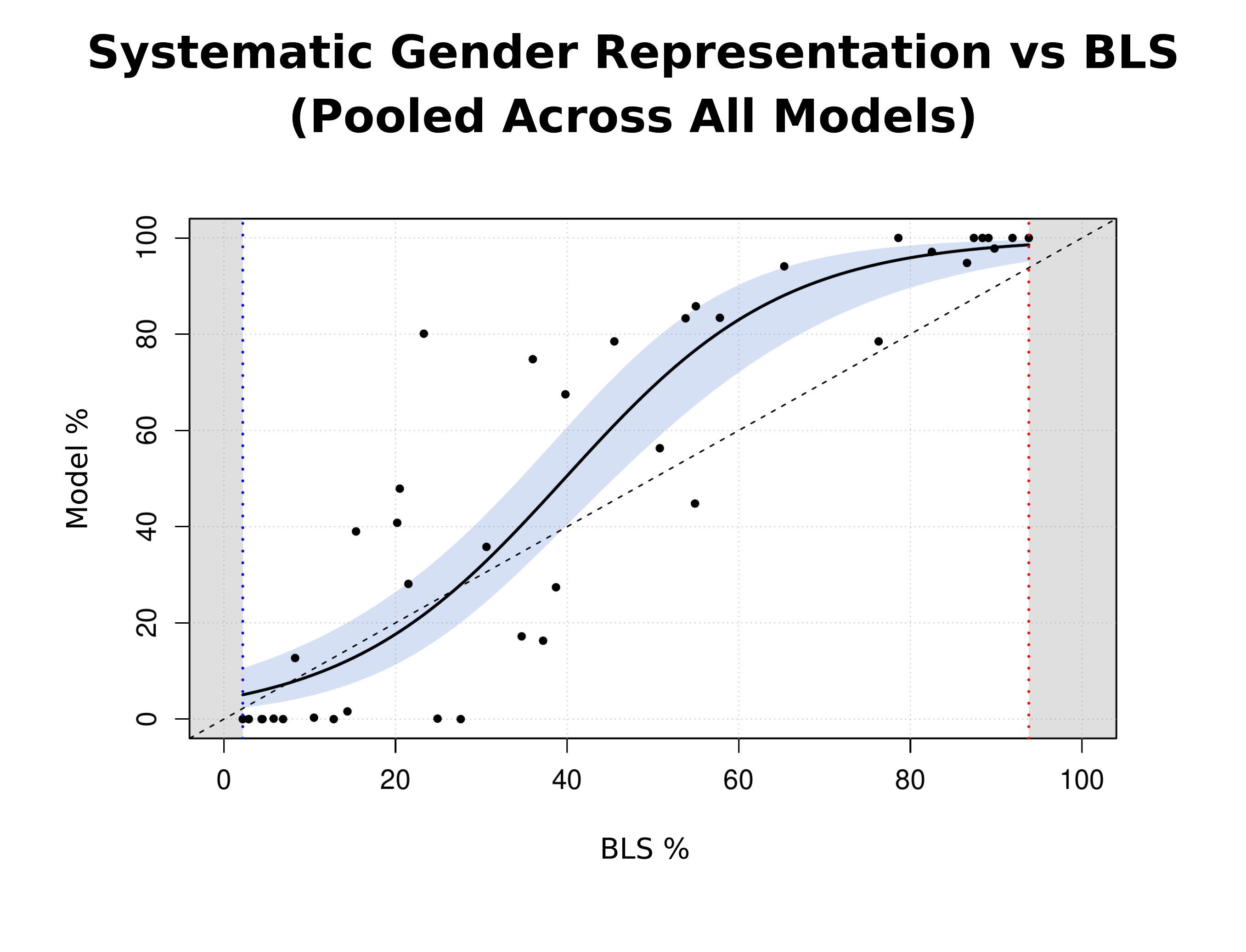}
  \vspace{-18pt}
  \caption{Systematic gender representation pooled across all models. Each point is an occupation, plotting BLS percentage of women against pooled model output. The diagonal line indicates parity. The fitted curve follows an S-shaped pattern, showing that models tend to exaggerate gender skews at both male- and female-dominated ends of the distribution. }
  \label{fig:rq2_gender}
  \Description{Scatterplot with BLS share of women on the x-axis and model share on the y-axis. The fitted line forms an S-shape, crossing near parity in the middle but bending away at both ends, indicating exaggeration of gender imbalances.}
\end{figure}


\subsubsection{Gender.}
\autoref{fig:rq2_gender} plots the share of women generated by models against BLS baselines, pooling across all models. The fitted curve follows an S-shaped pattern: models underpredict women in male-dominated occupations and overpredict them in female-dominated ones, amplifying existing gender imbalances.

\begin{figure}[H]
  \centering
  \includegraphics[width=\linewidth, scale=1.15]{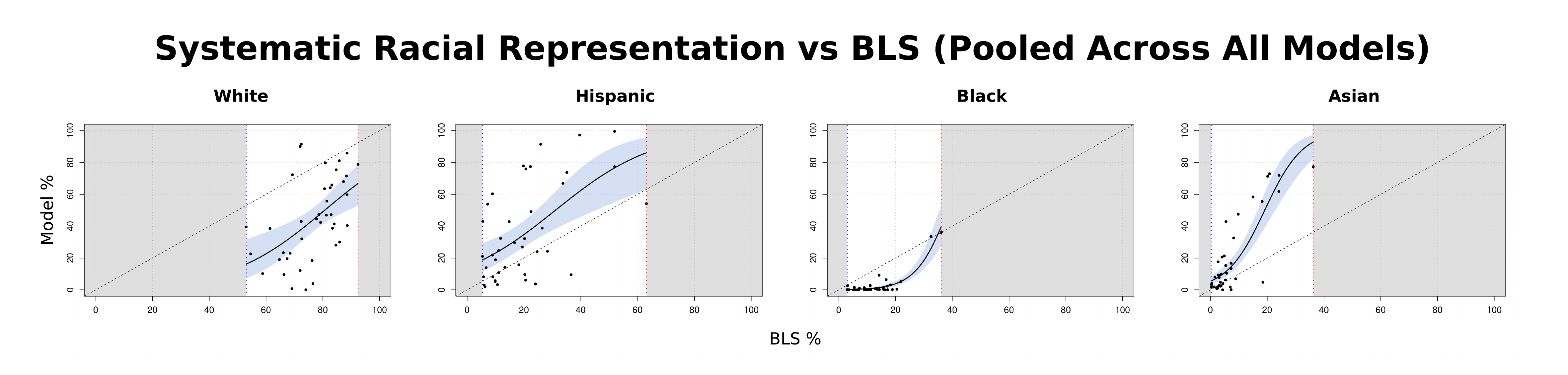}
  \vspace{-16pt}
  \caption{Systematic racial representation pooled across all models. Each point is an occupation, plotting BLS percentages of racial groups against pooled model outputs. The diagonal line indicates parity. The fitted curves show underrepresentation of Black and White workers, overrepresentation of Hispanic and Asian workers, exaggeration of existing skews for White, Hispanic, and Asian groups, and suppression of Black representation across most occupations rather than amplification.}
  \label{fig:rq2_race}
  \Description{Four panels plotting BLS share (x-axis) against model share (y-axis) for White, Hispanic, Black, and Asian workers. Curves fall below parity for Black and White groups and rise above parity for Hispanic and Asian groups. Slopes for White, Hispanic, and Asian groups bend away from the diagonal, indicating exaggeration of occupational skews, while the Black curve runs nearly flat, indicating suppression rather than amplification.}
\end{figure}

\autoref{tab:rq2_race} quantifies these patterns. At the median occupation (about 36\% women in BLS data), models underrepresent women by 6.1 percentage points. The positive slope deviation (+0.17) indicates stereotype exaggeration: the gap between models and BLS grows as occupations become more strongly segregated. The regression and S-curve converge on the same conclusion: LLMs exaggerate gender stereotypes rather than reducing them.

\begin{table}[H]
\centering
\caption{{Systematic Racial and Gender Representation (Pooled). Logistic regression estimates of systematic bias for racial and  gender representation pooled across 41 occupations. Over/underrepresentation is based on the intercept ($\alpha$), and stereotyping is based on the slope ($\beta$). Stars indicate FDR-adjusted significance (*$p < .05$, **$p < .01$, ***$p < .001$). For significant results, symbol codes (\symzero, \symplus, \symplusplus, \symplusplusplus, \symminus, \symminusminus, \symminusminusminus) summarize the direction and strength of effects. }}
\label{tab:rq2_race}
\begin{adjustbox}{max width=\linewidth}
\begin{tabular}{lrrccrrc}
\toprule
                                   & \multicolumn{3}{c}{\textbf{Systematic Shift}}                                                                                             && \multicolumn{3}{c}{\textbf{Stereotype Slope}}                                                                                       \\
\cmidrule{2-4} \cmidrule{6-8}
\multicolumn{1}{c}{\textbf{Group}} & \multicolumn{1}{c}{\textbf{$\alpha$ (logit)}} & \multicolumn{1}{c}{\textbf{Systematic shift (pp)}} & \multicolumn{1}{c}{\textbf{Interpretation}} && \multicolumn{1}{c}{\textbf{$\beta$ (logit)}} & \multicolumn{1}{c}{\textbf{Slope deviation}} & \multicolumn{1}{c}{\textbf{Interpretation}} \\
\midrule
White                              & \num{-0.14}    & \num{-32.1}***      & \symminusminusminus    && \num{9.65}    & +\num{1.13}***   & \symplusplus     \\
Hispanic                           & \num{-0.96}    & \num{9.4}***        & \symplusplus           && \num{6.93}    & +\num{0.43}***   & \symplus         \\
Black                              & \num{-5.81}    & \num{-10.8}***      & \symminusminusminus    && \num{21.21}   & \num{-0.89}***   & \symminusminus   \\
Asian                              & \num{-2.47}    & \num{3.4}***        & \symplusplus           && \num{18.89}   & +\num{2.10}***   & \symplusplusplus \\
\cmidrule(lr){1-8}
Women          & \num{-0.85}              & \num{-6.1}***                     & \symminusminus                     && \num{13.26}              & +\num{0.17}***                 & \symplus                     \\
\bottomrule
\end{tabular}
\end{adjustbox}
\end{table}

\subsubsection{Race.} 
\autoref{fig:rq2_race} and \autoref{tab:rq2_race} extend this analysis to racial categories. Here, the distortions vary across groups. Black workers are consistently underrepresented: the regression shows a large negative shift (–10.8 pp) and a slope below parity (–0.89), indicating stereotype reduction through near-erasure at typical occupational levels, with presence surfacing only in a handful of outlier occupations (such as \textit{Security Guard} and \textit{Bus Driver}).

White workers are also underrepresented overall (–32.1 pp), but with a positive slope (+1.13), meaning that their share is exaggerated in occupations where they are already more common. By contrast, Hispanic (+9.4 pp) and Asian workers (+3.4 pp) are systematically overrepresented. For both groups, positive slope deviations (Hispanic +0.43; Asian +2.10) indicate stereotype exaggeration, with the gap between models and BLS widening in occupations where these groups already have higher representation.

The shift/exaggeration decomposition captures two coexisting forms of bias: uniform demographic shifts ($\alpha$) and occupation-level stereotype amplification ($\beta$). Most groups exhibit both, with over- or underrepresentation paired with amplification of existing skews. Black workers are the exception: the negative slope deviation reflects suppression across the median range rather than amplification.

\subsection{RQ3: Cross-Model Comparisons of Systematic Bias}
A core question is whether the observed distortions are unique artifacts of specific models or evidence of a shared structural bias across the field. By comparing GPT-4, Gemini, DeepSeek, and Mistral (models developed under different institutional and cultural regimes), we find further evidence of structural convergence in their representational outputs.


\begin{table}[H]
    \centering
    \caption{Systematic Gender Representation (By Model). Logistic regression estimates of systematic bias for gender representation, disaggregated by model. Notation as in Table~\ref{tab:rq2_race}.}
    \label{tab:rq3_gender}
\begin{adjustbox}{max width=\linewidth}
\begin{tabular}{llrrccrrc}
\toprule
                                       &                                    & \multicolumn{3}{c}{\textbf{Systematic Shift}}                                                                                             && \multicolumn{3}{c}{\textbf{Stereotype Slope}}                                                                                       \\
\cmidrule{3-5} \cmidrule{7-9}
\multicolumn{1}{c}{\textbf{Model}} & \multicolumn{1}{c}{\textbf{Group}} & \multicolumn{1}{c}{\textbf{$\alpha$ (logit)}} & \multicolumn{1}{c}{\textbf{Systematic shift (pp)}} & \multicolumn{1}{c}{\textbf{Interpretation}} && \multicolumn{1}{c}{\textbf{$\beta$ (logit)}} & \multicolumn{1}{c}{\textbf{Slope deviation}} & \multicolumn{1}{c}{\textbf{Interpretation}} \\
\midrule
GPT-4       & \multirow{4}{*}{Women}    & \num{-0.13}  & \num{10.7}***  & \symplusplusplus  && \num{14.91}   & +\num{0.17}***  & \symplus    \\
Gemini        &     & \num{-0.93}  & \num{-7.6}***  & \symminusminus  && \num{6.81 }   & +\num{0.05}***  & \symzero    \\
Mistral       &     & \num{-0.29}  & \num{6.9}***   & \symplusplus    && \num{19.41}   & +\num{0.18}***  & \symplus    \\
DeepSeek      &     & \num{-0.22}  & \num{9.1}***   & \symplusplus    && \num{27.61}   & +\num{0.19}***  & \symplus    \\
\bottomrule
\end{tabular}
\end{adjustbox}
\end{table}

\subsubsection{Gender.}
Table~\ref{tab:rq3_gender} shows broadly similar patterns across models. GPT-4, Mistral, and DeepSeek lift women relative to BLS (+10.7~pp, +6.9~pp, and +9.1~pp), while Gemini exhibits a downward offset ($-7.6$~pp). Stereotype slope deviations are positive but small across all four models, with Gemini closest to zero. Cross-model differences for gender are driven by variation in the systematic shift rather than qualitative differences in stereotype encoding.


\subsubsection{Race.}
Full by-model racial regression estimates (Table~\ref{tab:rq3_race}) show that racial and ethnic representation exhibits several similarities across models. All four models underrepresent Black workers on average, with nearly identical median shifts of approximately $-11$~pp across providers. White workers are also systematically underrepresented, though with greater variation in magnitude (e.g., Gemini $-44.3$~pp; Mistral $-34.6$~pp; GPT-4 $-23.7$~pp; DeepSeek $-17.7$~pp). In contrast, Hispanic and Asian workers are generally overrepresented across models.

Stereotype slope deviations are positive for Hispanic and Asian workers, indicating modest exaggeration of existing occupational skews, while slopes for Black workers are weak or negative, reflecting suppression across a broad range of occupations rather than concentration in already skewed roles. Overall, cross-model differences for race and ethnicity are driven primarily by variation in the magnitude of systematic shifts and exaggeration, rather than by qualitative differences in which groups are over- or underrepresented.

\subsubsection{Salary representation.}
\rev{In addition to demographic attributes, we compared model-generated salaries to BLS median annual earnings for each occupation. All four models track BLS salary rankings moderately well (Spearman $\rho = 0.73$--$0.80$), though all substantially overestimate salaries for high-prestige occupations such as CEO. When examining salary by gender, BLS suppresses earnings estimates for demographic subgroups where the estimated full-time workforce falls below 50{,}000 workers~} \cite{bls2023cpsaat39}, \rev{limiting gender comparisons to 25 of 41 occupations. We further restrict to cases where a given model generates personas of both men and women, yielding between 11 and 24 occupation--model pairs depending on the model. BLS data shows a median gender wage gap of \$7{,}592 favoring men across these occupations. Three of four models compress this gap to a median of \$0, assigning identical salaries to men and women in a given occupation; Gemini shows a small residual gap (median \$750). For example, BLS reports that men in software developer roles earn approximately \$14{,}000 more annually than women; three of four models generate identical salaries for men and women in this occupation. Models thus largely treat salary as a property of the occupation rather than conditioning on demographic attributes, compressing the real-world gender pay disparity. Full salary comparisons, including occupation-level breakdowns, are reported in Appendix~} \ref{app:salary} \rev{.}

\section{Discussion}

\subsection{Interpreting systematic bias}



Interpreted through our five-pattern framework, the results show that representational bias in LLM-generated personas is structured distortion, not random noise. Two mechanisms recur across models: \emph{systematic shifts in visibility} and \emph{stereotype-dependent concentration}. Gender representation follows the S-shaped exaggeration pattern observed in prior audits of image search and generative media \cite{Guilbeault2024-ej,Kay2015-yl}. For race, Hispanic and Asian representation shows stereotype exaggeration, while Black representation exhibits systematic suppression across most occupations rather than amplification in already-skewed roles.

This latter pattern is particularly notable. Apparent ``stereotype reduction'' for Black workers does not reflect balanced representation, but near-erasure across the median occupational range, with presence concentrated in a small number of stereotyped roles. Similar dynamics have been documented for marginalized identities in other algorithmic systems \cite{Scheuerman2019-en,Scheuerman2020-bk}, suggesting that suppression can be as consequential a representational harm as exaggeration.

\rev{The concentration analysis reveals that these patterns are especially pronounced for gender, where most model--occupation pairs reflect mode selection rather than distributional sampling. Racial generation is more variable across models, but still consistently more concentrated than real-world data. Because three of the four audited models converge on similar demographic defaults, practitioners encounter what amounts to algorithmic monoculture~} \cite{kleinberg2021} \rev{in generated personas, with identical demographic assignments regardless of which model is used. This compounds the narrowing of representational diversity that Bommasani et al.\ identify as a core risk of foundation model homogenization~} \cite{bommasani2022}.

\rev{The decomposition also provides a framework for comparing across system types. Prior audits of image search~} \cite{Kay2015-yl,Metaxa2021-ty} \rev{identified uniform underrepresentation alongside modest stereotype exaggeration. Our LLM results reveal a structurally different pattern: overrepresentation of some groups combined with strong exaggeration. These diagnostically distinct bias regimes call for different interventions, a distinction that aggregate measures would obscure.}

\subsection{Cross-model audits reveal structural rather than model-specific bias}
\rev{The consistency of bias patterns across models with different institutional origins (RQ3) is consistent with shared upstream influences rather than model-specific design choices. This aligns with prior work showing that large language models trained on web-scale corpora reproduce the social patterns embedded in their training data} \cite{Bender2021-lb} \rev{and with audits of generative image systems finding the same directional biases across different architectures and providers} \cite{Bianchi2023-zt}. \rev{Common training data sources, widely circulating cultural associations between occupations and demographics} \cite{Bolukbasi2016-bi} \rev{and representational conventions in web-scale text are plausible contributing factors. While post-training interventions may modulate the intensity of bias, they appear insufficient to alter its underlying structure.}

\rev{Gemini's partial shift toward distributional generation for race suggests that post-training interventions can alter the generation regime, not just the direction of bias. However, Gemini remains predominantly modal for gender and collapses to stereotyped defaults for high-association occupations such as housekeeper and cook, indicating that diversity interventions address some dimensions of representation while leaving others structurally unchanged. This underscores the importance of cross-model audits: single-model analyses risk conflating structural patterns with idiosyncratic design choices, or vice versa.}

\subsection{Undesirable and ambiguous representativeness}

A central finding of RQ1 is that representational distortion is not inherently good or bad; its implications depend on context, magnitude, and historical meaning. Some patterns clearly reinforce harmful stereotypes, such as the near-exclusive portrayal of housekeepers as Hispanic across all four models, echoing cultural tropes that associate Hispanic identity with low-status labor \cite{del-Rio-Gabiola2013-cb,Yemane2021-bs}. Yet not all deviations are unambiguously harmful: women are frequently overrepresented in high-status professional roles where they remain underrepresented in reality, a pattern that could be interpreted as aspirational. These cases illustrate that deviation from descriptive benchmarks alone is insufficient to determine harm \cite{Baumer2017-ol,Metaxa2021-ty}, suggesting the need for evaluation frameworks that make representational choices explicit and adjustable.

\subsection{Limitations}

Our analysis has several important limitations. First, we benchmarked model outputs against U.S.\ Bureau of Labor Statistics (BLS) data. This offers a consistent and widely used point of comparison, but it is not a normative standard. BLS distributions themselves reflect historical inequities in the U.S.\ labor market (e.g., \cite{Blau2017-mi, Bertrand2004-bs}), so parity with BLS does not necessarily equate to fairness, and deviations from BLS do not automatically signal harm. The BLS dataset also excludes non-binary identities and generalizes racial/ethnic groups into broad categories, limiting the robustness of our analyses. Accordingly, following prior work that highlights how descriptive accuracy can naturalize existing inequities \cite{Kay2015-yl, Metaxa2021-ty}, we treat our results as descriptive evidence rather than prescriptive judgments about fairness.


Our results are sensitive to prompt and parameter choices. We used a single, fixed prompt and default API settings to maintain comparability, but prior work shows that even small paraphrases can shift representational outcomes \cite{Chen2025-bm}. The convergence across four models suggests these patterns reflect biases in shared training data, though we cannot rule out shared prompt sensitivities. A natural extension would vary prompt wording or sampling temperature, but some audited model versions have since been retired by their providers (see Future Work).



It is also possible that LLMs interpret some occupation labels differently than BLS categories. However, the highest-deviation occupations (e.g., housekeeper, security guard) correspond to unambiguous labels with clear stereotypical associations rather than taxonomic edge cases. As a partial robustness check, our trimmed regression specification (which down-weights the most extreme occupation-level outliers) yields the same qualitative conclusions, indicating that the patterns reported here are not driven by a small number of ambiguous labels (see Appendix~\ref{app:regression_tables}).

Finally, our primary analyses focused on structured demographic attributes (gender and race/ethnicity). Generated names, motivations, biographies, and other narrative fields likely encode additional stereotypes that warrant systematic analysis beyond the scope of this study.

\subsection{Future work} 
Promising directions include deepening mixed-race and intersectional analyses, extending audits to non-binary identities, examining representational bias in narrative attributes (e.g., motivations or biographies), and applying benchmarks from other national labor markets. In addition, adopting multi-prompt and multi-parameter strategies \cite{Chen2025-bm} would help assess whether the patterns observed here are robust across elicitation conditions or primarily reflect shared prompt sensitivities across current model architectures. A particularly informative direction would be testing whether simple prompt modifications (e.g., ``a typical'' versus ``a random'' engineer) or temperature adjustments can shift models from modal to distributional generation regimes, clarifying both the source of the modal default and the leverage available to practitioners to push back against it.

\section{Conclusion}

\rev{Our audit of over 1.5 million generative AI occupational personas across 41 U.S. occupations finds that gender generation is predominantly modal, with models assigning a single dominant gender per occupation, while racial generation is more variable across models but consistently more concentrated than real-world data. A shift/exaggeration decomposition reveals systematic overrepresentation for women, Asian, and Hispanic workers, and underrepresentation for White and Black workers, with stereotypes exaggerated for most groups while being reduced for Black workers through near erasure at typical occupational levels.} \rev{These patterns recur across four models with different institutional origins, pointing to shared structural sources of bias. The shift/exaggeration decomposition introduced here offers a reusable diagnostic for future audits, separating uniform demographic shifts from occupation-specific stereotype amplification in ways that aggregate bias measures cannot. Our cross-model comparison further demonstrates that single-model audits risk conflating structural patterns with idiosyncratic design choices, underscoring the need for multi-model evaluation as a standard auditing practice. As LLM-generated personas become embedded in product design, hiring simulations, and synthetic data pipelines, continued work is needed to surface and make adjustable the demographic assumptions these systems encode.}

\section*{Generative AI Usage Statement}

Per ACM’s Policy on Authorship \cite{ACMAuthorshipPolicy2023}, generative AI tools were used in two ways: (1) to produce the synthetic occupational personas constituting the dataset, and (2) for language editing, script refactoring, and formatting assistance (GPT-4, Sonnet 4.6, and GPT-5). All study design, analyses, interpretations, and claims are those of the human authors.

\begin{acks}
We thank Ulrik Lyngs for feedback on figures and discussion and Sean Munson for feedback on the study design. This work was supported by the Stitt Fellowship Program at the Human-Computer Interaction Lab at Santa Clara University and the Faculty Hackworth Grant from the Markkula Center for Applied Ethics at Santa Clara University.
\end{acks}

\bibliographystyle{ACM-Reference-Format}

\bibliography{ref}
\begin{DIFnomarkup}
\appendix
\clearpage

\section{Per-Model Regression Plots}
\label{app:regression_plots}

\begin{figure}[htbp]
  \centering
  \includegraphics[width=\textwidth, scale=0.75]{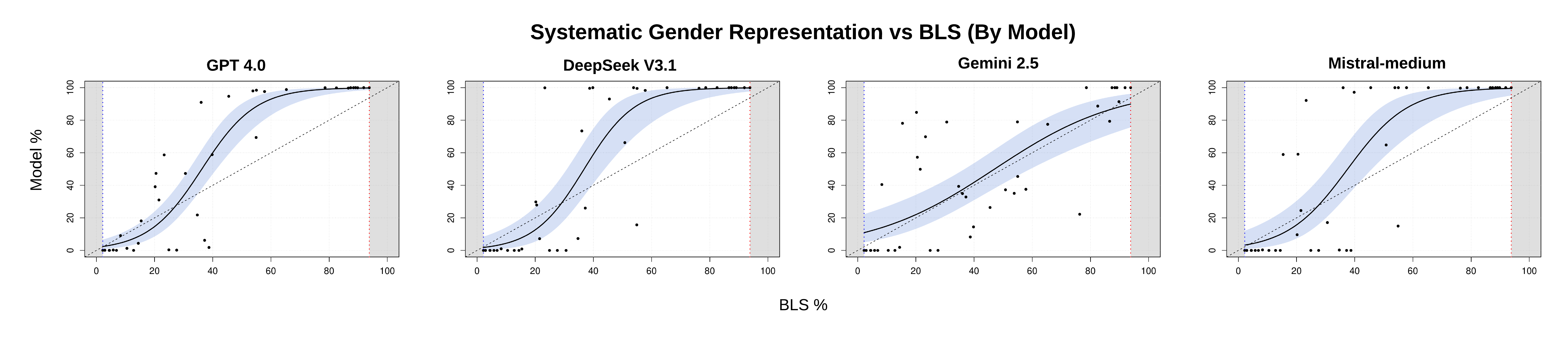}
  \caption{Systematic gender representation by model. Each panel plots BLS percentage of women (x-axis) against model-generated percentage (y-axis) for one model. The diagonal line indicates parity. All four models exhibit S-shaped exaggeration, amplifying gender skews at both ends of the distribution. Gemini shows the weakest exaggeration and a slight downward shift, while GPT-4, Mistral, and DeepSeek show upward shifts indicating overrepresentation of women.}
  \label{fig:app_gender_by_model}
  \Description{Four scatterplots, one per model, each showing BLS percentage of women on the x-axis and model percentage on the y-axis. Fitted S-curves cross near the middle but bend away from the parity diagonal at both ends, indicating stereotype exaggeration. Gemini's curve sits below parity while the other three sit above.}
\end{figure}

\begin{figure}[htbp]
  \centering
  \includegraphics[width=\textwidth, scale=0.75]{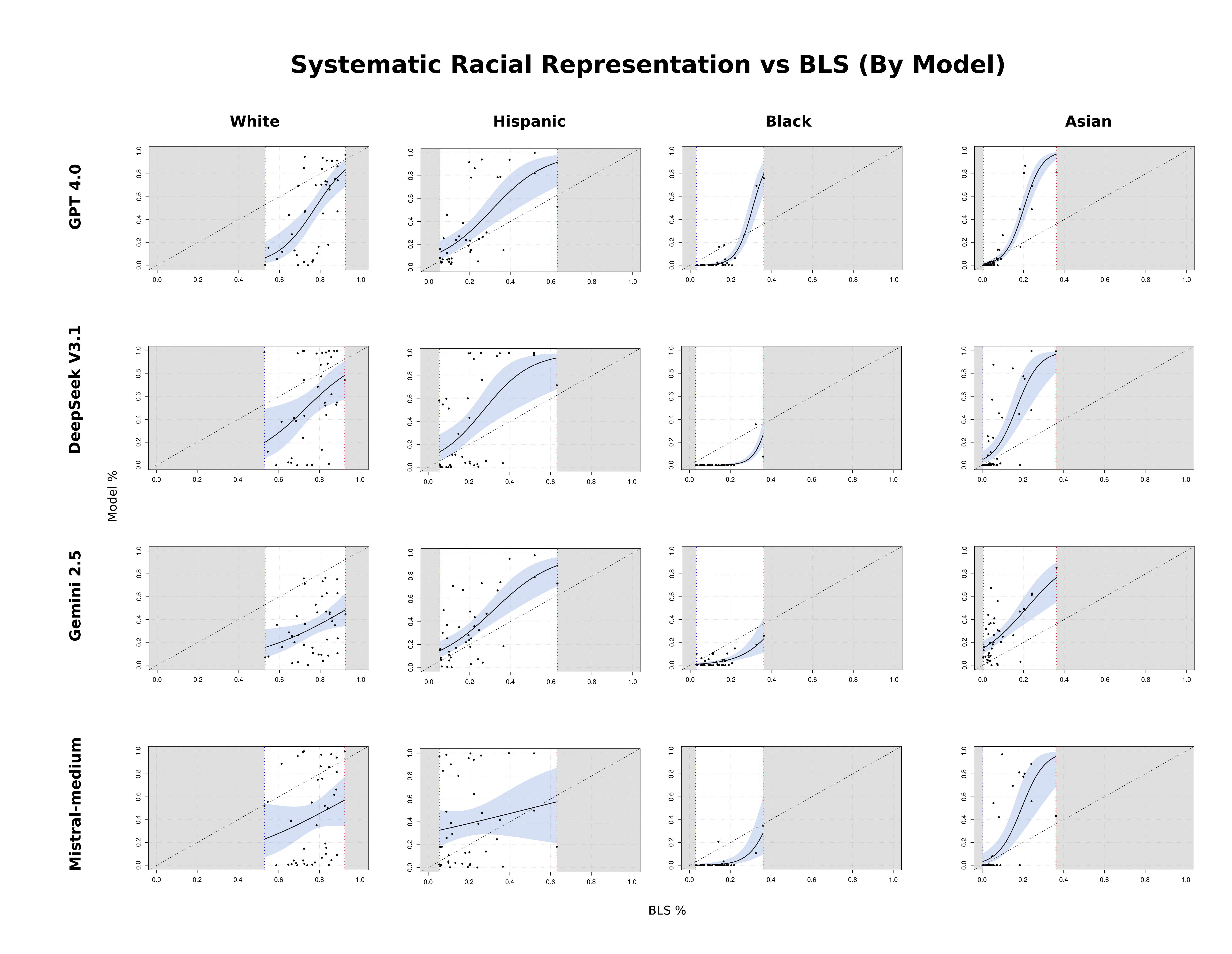}
  \caption{Systematic racial representation by model. Each row shows one racial/ethnic group (White, Hispanic, Black, Asian) and each column shows one model. The diagonal line indicates parity. All four models underrepresent White and Black workers and overrepresent Hispanic and Asian workers. Black workers show a distinctive pattern of near-erasure (slope below parity), while Hispanic and Asian workers show exaggeration of existing occupational skews.}
  \label{fig:app_race_by_model}
  \Description{Grid of 16 scatterplots (4 racial groups by 4 models) comparing BLS share to model share. White and Black panels show points mostly below the diagonal; Hispanic and Asian panels show points above. Fitted curves for Black workers are notably flat, indicating suppression across the occupational range.}
\end{figure}

\clearpage
\section{Full Regression Tables}
\label{app:regression_tables}

\begin{table}[H]
    \centering
    \caption{Systematic Racial Representation (By Model). Logistic regression estimates of systematic bias for racial representation, disaggregated by model. Notation as in Table~\ref{tab:rq2_race}.}
    \label{tab:rq3_race}
\begin{adjustbox}{max width=\linewidth}
\begin{tabular}{llrrccrrc}
\toprule
                                   &                                    & \multicolumn{3}{c}{\textbf{Systematic Shift}}               & & \multicolumn{3}{c}{\textbf{Stereotype Slope}} \\
\cmidrule{3-5} \cmidrule{7-9}
\multicolumn{1}{c}{\textbf{Model}} & \multicolumn{1}{c}{\textbf{Group}} & \multicolumn{1}{c}{\textbf{$\alpha$ (logit)}} & \multicolumn{1}{c}{\textbf{Systematic shift (pp)}} & \multicolumn{1}{c}{\textbf{Interpretation}} && \multicolumn{1}{c}{\textbf{$\beta$ (logit)}} & \multicolumn{1}{c}{\textbf{Slope deviation}} & \multicolumn{1}{c}{\textbf{Interpretation}} \\
\midrule
\multirow{4}{*}{GPT-4}
 & White     & \num{0.20}  & \num{-23.7}*** & \symminusminusminus   && \num{11.65} & +\num{1.52}*** & \symplusplusplus \\
 & Hispanic  & \num{-1.21} & \num{4.7}***   & \symplusplus          && \num{10.24} & +\num{0.91}*** & \symplusplus \\
 & Black     & \num{-6.45} & \num{-10.9}*** & \symminusminusminus   && \num{33.07} & \num{-0.85}*** & \symminusminus \\
 & Asian     & \num{-4.14} & \num{-2.8}***  & \symminus             && \num{27.98} & +\num{2.03}*** & \symplusplusplus \\
\midrule
\multirow{4}{*}{Gemini}
 & White     & \num{-0.65} & \num{-44.3}*** & \symminusminusminus   && \num{4.35}  & \num{-0.06}*** & \symzero \\
 & Hispanic  & \num{-1.10} & \num{6.8}***   & \symplusplus          && \num{7.92}  & +\num{0.56}*** & \symplusplus \\
 & Black     & \num{-5.71} & \num{-10.8}*** & \symminusminusminus   && \num{18.70} & \num{-0.90}*** & \symminusminus \\
 & Asian     & \num{-1.58} & \num{12.6}***  & \symplusplusplus      && \num{10.41} & +\num{1.05}*** & \symplusplus \\
\midrule
\multirow{4}{*}{Mistral}
 & White     & \num{-0.24} & \num{-34.6}*** & \symminusminusminus   && \num{5.83}  & +\num{0.36}*** & \symplus \\
 & Hispanic  & \num{-0.81} & \num{12.5}***  & \symplusplusplus      && \num{2.45}  & \num{-0.47}*** & \symminus \\
 & Black     & \num{-10.81}& \num{-11.1}*** & \symminusminusminus   && \num{40.70} & \num{-1.00}*** & \symminusminus \\
 & Asian     & \num{-5.89} & \num{-4.1}***  & \symminusminus        && \num{42.07} & +\num{2.67}*** & \symplusplusplus \\
\midrule
\multirow{4}{*}{DeepSeek}
 & White     & \num{0.45}  & \num{-17.7}*** & \symminusminusminus   && \num{10.13} & +\num{1.24}*** & \symplusplus \\
 & Hispanic  & \num{-1.49} & \num{-0.5}***  & \symzero              && \num{27.65} & +\num{2.16}*** & \symplusplusplus \\
 & Black     & \num{-10.91}& \num{-10.6}*** & \symminusminusminus   && \num{33.23} & \num{-1.00}*** & \symminusminus \\
 & Asian     & \num{-5.04} & \num{-3.7}***  & \symminusminus        && \num{37.43} & +\num{2.80}*** & \symplusplusplus \\
\bottomrule
\end{tabular}
\end{adjustbox}
\end{table}


\clearpage
\begin{table}[H]
\centering
\caption{Full Regression Coefficients --- GPT-4. Raw logit coefficients, standard errors, and $p$-values for the intercept ($\alpha$) and slope ($\beta$) from logistic regression models of representation, by demographic group and estimation method. $p$-values reported unadjusted and with Benjamini--Hochberg FDR correction.}
\label{tab:coefs_chatgpt}
\begin{adjustbox}{max width=\linewidth}
\begin{tabular}{llrrrrrrrrr}
\toprule
& & \multicolumn{4}{c}{\textbf{Intercept ($\alpha$)}} & & \multicolumn{4}{c}{\textbf{Slope ($\beta$)}} \\
\cmidrule{3-6} \cmidrule{8-11}
\multicolumn{1}{c}{\textbf{Group}} &
\multicolumn{1}{c}{\textbf{Method}} &
\multicolumn{1}{c}{\textbf{Coef.}} &
\multicolumn{1}{c}{\textbf{SE}} &
\multicolumn{1}{c}{\textbf{$p$}} &
\multicolumn{1}{c}{\textbf{$p_{\text{FDR}}$}} & &
\multicolumn{1}{c}{\textbf{Coef.}} &
\multicolumn{1}{c}{\textbf{SE}} &
\multicolumn{1}{c}{\textbf{$p$}} &
\multicolumn{1}{c}{\textbf{$p_{\text{FDR}}$}} \\
\midrule
\multirow{3}{*}{Asian}
   & Raw      & $-3.438$ & $0.306$ & $2.61\times10^{-29}$ & $2.61\times10^{-28}$ && $22.515$ & $2.413$ & $4.86\times10^{-19}$ & $9.72\times10^{-18}$ \\
 & Robust   & $-4.141$ & $0.014$ & $<0.001$             & $<0.001$             && $27.977$ & $0.101$ & $<0.001$             & $<0.001$             \\
 & Trimmed5 & $-3.631$ & $0.305$ & $1.12\times10^{-32}$ & $1.12\times10^{-31}$ && $25.681$ & $2.599$ & $2.14\times10^{-21}$ & $4.28\times10^{-20}$ \\
\cmidrule{1-11}
\multirow{3}{*}{Black}
   & Raw      & $-4.963$ & $0.456$ & $1.57\times10^{-27}$ & $1.04\times10^{-26}$ && $25.457$ & $2.914$ & $4.77\times10^{-17}$ & $4.77\times10^{-16}$ \\
 & Robust   & $-6.448$ & $0.031$ & $<0.001$             & $<0.001$             && $33.068$ & $0.169$ & $<0.001$             & $<0.001$             \\
 & Trimmed5 & $-4.775$ & $0.565$ & $2.90\times10^{-17}$ & $1.45\times10^{-16}$ && $20.823$ & $8.567$ & $2.07\times10^{-2}$  & $2.95\times10^{-2}$  \\
\cmidrule{1-11}
\multirow{3}{*}{Hispanic}
   & Raw      & $-0.857$ & $0.218$ & $8.62\times10^{-5}$  & $1.57\times10^{-4}$  && $7.718$  & $1.845$ & $2.72\times10^{-4}$  & $3.88\times10^{-4}$  \\
 & Robust   & $-1.215$ & $0.004$ & $<0.001$             & $<0.001$             && $10.235$ & $0.038$ & $<0.001$             & $<0.001$             \\
 & Trimmed5 & $-0.847$ & $0.217$ & $9.83\times10^{-5}$  & $1.97\times10^{-4}$  && $9.992$  & $2.166$ & $3.31\times10^{-5}$  & $1.32\times10^{-4}$  \\
\cmidrule{1-11}
\multirow{3}{*}{White}
   & Raw      & $0.039$  & $0.202$ & $0.847$              & $0.847$              && $9.937$  & $2.436$ & $2.44\times10^{-4}$  & $3.75\times10^{-4}$  \\
 & Robust   & $0.196$  & $0.004$ & $<0.001$             & $<0.001$             && $11.650$ & $0.045$ & $<0.001$             & $<0.001$             \\
 & Trimmed5 & $0.014$  & $0.215$ & $0.949$              & $0.949$              && $9.332$  & $2.865$ & $3.64\times10^{-3}$  & $6.62\times10^{-3}$  \\
\cmidrule{1-11}
\multirow{3}{*}{Women}
   & Raw      & $-0.137$ & $0.226$ & $0.545$              & $0.641$              && $10.059$ & $1.384$ & $5.98\times10^{-11}$ & $2.99\times10^{-10}$ \\
 & Robust   & $-0.132$ & $0.006$ & $2.01\times10^{-99}$ & $2.23\times10^{-99}$ && $14.913$ & $0.053$ & $<0.001$             & $<0.001$             \\
 & Trimmed5 & $-0.134$ & $0.233$ & $0.565$              & $0.665$              && $9.929$  & $1.447$ & $6.78\times10^{-10}$ & $6.78\times10^{-9}$  \\
\bottomrule
\end{tabular}
\end{adjustbox}
\end{table}

\begin{table}[H]
\centering
\caption{Full Regression Coefficients --- Gemini. Raw logit coefficients, standard errors, and $p$-values for the intercept ($\alpha$) and slope ($\beta$) from logistic regression models of representation, by demographic group and estimation method. $p$-values reported unadjusted and with Benjamini--Hochberg FDR correction.}
\label{tab:coefs_gemini}
\begin{adjustbox}{max width=\linewidth}
\begin{tabular}{llrrrrrrrrr}
\toprule
& & \multicolumn{4}{c}{\textbf{Intercept ($\alpha$)}} & & \multicolumn{4}{c}{\textbf{Slope ($\beta$)}} \\
\cmidrule{3-6} \cmidrule{8-11}
\multicolumn{1}{c}{\textbf{Group}} &
\multicolumn{1}{c}{\textbf{Method}} &
\multicolumn{1}{c}{\textbf{Coef.}} &
\multicolumn{1}{c}{\textbf{SE}} &
\multicolumn{1}{c}{\textbf{$p$}} &
\multicolumn{1}{c}{\textbf{$p_{\text{FDR}}$}} & &
\multicolumn{1}{c}{\textbf{Coef.}} &
\multicolumn{1}{c}{\textbf{SE}} &
\multicolumn{1}{c}{\textbf{$p$}} &
\multicolumn{1}{c}{\textbf{$p_{\text{FDR}}$}} \\
\midrule
\multirow{3}{*}{Asian}
   & Raw      & $-1.397$ & $0.171$ & $3.04\times10^{-16}$ & $1.21\times10^{-15}$ && $8.155$  & $1.778$ & $5.71\times10^{-5}$  & $1.14\times10^{-4}$  \\
 & Robust   & $-1.584$ & $0.005$ & $<0.001$             & $<0.001$             && $10.409$ & $0.049$ & $<0.001$             & $<0.001$             \\
 & Trimmed5 & $-1.351$ & $0.182$ & $1.03\times10^{-13}$ & $4.11\times10^{-13}$ && $6.639$  & $2.410$ & $1.93\times10^{-2}$  & $2.95\times10^{-2}$  \\
\cmidrule{1-11}
\multirow{3}{*}{Black}
   & Raw      & $-3.808$ & $0.290$ & $1.79\times10^{-39}$ & $3.59\times10^{-38}$ && $10.446$ & $2.256$ & $2.84\times10^{-5}$  & $7.09\times10^{-5}$  \\
 & Robust   & $-5.714$ & $0.028$ & $<0.001$             & $<0.001$             && $18.705$ & $0.146$ & $<0.001$             & $<0.001$             \\
 & Trimmed5 & $-3.987$ & $0.315$ & $8.46\times10^{-37}$ & $1.69\times10^{-35}$ && $11.743$ & $5.475$ & $4.98\times10^{-2}$  & $6.63\times10^{-2}$  \\
\cmidrule{1-11}
\multirow{3}{*}{Hispanic}
   & Raw      & $-0.904$ & $0.178$ & $4.01\times10^{-7}$  & $8.92\times10^{-7}$  && $6.703$  & $1.408$ & $5.13\times10^{-5}$  & $1.14\times10^{-4}$  \\
 & Robust   & $-1.099$ & $0.004$ & $<0.001$             & $<0.001$             && $7.922$  & $0.033$ & $<0.001$             & $<0.001$             \\
 & Trimmed5 & $-0.905$ & $0.192$ & $2.49\times10^{-6}$  & $5.54\times10^{-6}$  && $7.478$  & $1.700$ & $1.39\times10^{-4}$  & $3.96\times10^{-4}$  \\
\cmidrule{1-11}
\multirow{3}{*}{White}
   & Raw      & $-0.630$ & $0.159$ & $7.23\times10^{-5}$  & $1.45\times10^{-4}$  && $4.113$  & $1.748$ & $7.50\times10^{-2}$  & $8.34\times10^{-2}$  \\
 & Robust   & $-0.648$ & $0.003$ & $<0.001$             & $<0.001$             && $4.347$  & $0.039$ & $<0.001$             & $<0.001$             \\
 & Trimmed5 & $-0.580$ & $0.169$ & $6.12\times10^{-4}$  & $1.11\times10^{-3}$  && $3.825$  & $2.157$ & $0.190$              & $0.224$              \\
\cmidrule{1-11}
\multirow{3}{*}{Women}
   & Raw      & $-0.517$ & $0.247$ & $3.63\times10^{-2}$  & $5.58\times10^{-2}$  && $4.680$  & $0.920$ & $6.33\times10^{-5}$  & $1.15\times10^{-4}$  \\
 & Robust   & $-0.926$ & $0.005$ & $<0.001$             & $<0.001$             && $6.809$  & $0.020$ & $<0.001$             & $<0.001$             \\
 & Trimmed5 & $-0.516$ & $0.248$ & $3.72\times10^{-2}$  & $5.73\times10^{-2}$  && $4.321$  & $0.953$ & $4.91\times10^{-4}$  & $1.22\times10^{-3}$  \\
\bottomrule
\end{tabular}
\end{adjustbox}
\end{table}

\begin{table}[H]
\centering
\caption{Full Regression Coefficients --- Mistral. Raw logit coefficients, standard errors, and $p$-values for the intercept ($\alpha$) and slope ($\beta$) from logistic regression models of representation, by demographic group and estimation method. $p$-values reported unadjusted and with Benjamini--Hochberg FDR correction.}
\label{tab:coefs_mistral}
\begin{adjustbox}{max width=\linewidth}
\begin{tabular}{llrrrrrrrrr}
\toprule
& & \multicolumn{4}{c}{\textbf{Intercept ($\alpha$)}} & & \multicolumn{4}{c}{\textbf{Slope ($\beta$)}} \\
\cmidrule{3-6} \cmidrule{8-11}
\multicolumn{1}{c}{\textbf{Group}} &
\multicolumn{1}{c}{\textbf{Method}} &
\multicolumn{1}{c}{\textbf{Coef.}} &
\multicolumn{1}{c}{\textbf{SE}} &
\multicolumn{1}{c}{\textbf{$p$}} &
\multicolumn{1}{c}{\textbf{$p_{\text{FDR}}$}} & &
\multicolumn{1}{c}{\textbf{Coef.}} &
\multicolumn{1}{c}{\textbf{SE}} &
\multicolumn{1}{c}{\textbf{$p$}} &
\multicolumn{1}{c}{\textbf{$p_{\text{FDR}}$}} \\
\midrule
\multirow{3}{*}{Asian}
   & Raw      & $-2.626$ & $0.486$ & $6.35\times10^{-8}$  & $1.59\times10^{-7}$  && $17.606$ & $4.348$ & $1.34\times10^{-4}$  & $2.23\times10^{-4}$  \\
 & Robust   & $-5.887$ & $0.028$ & $<0.001$             & $<0.001$             && $42.070$ & $0.191$ & $<0.001$             & $<0.001$             \\
 & Trimmed5 & $-2.988$ & $0.513$ & $5.67\times10^{-9}$  & $1.89\times10^{-8}$  && $26.439$ & $5.381$ & $2.27\times10^{-6}$  & $1.52\times10^{-5}$  \\
\cmidrule{1-11}
\multirow{3}{*}{Black}
   & Raw      & $-5.397$ & $0.877$ & $7.58\times10^{-10}$ & $2.53\times10^{-9}$  && $17.870$ & $4.793$ & $4.32\times10^{-4}$  & $5.76\times10^{-4}$  \\
 & Robust   & $-10.812$ & $0.165$ & $<0.001$            & $<0.001$             && $40.704$ & $0.693$ & $<0.001$             & $<0.001$             \\
 & Trimmed5 & $-5.203$ & $0.936$ & $2.70\times10^{-8}$  & $6.75\times10^{-8}$  && $13.290$ & $15.746$ & $0.435$             & $0.435$              \\
\cmidrule{1-11}
\multirow{3}{*}{Hispanic}
   & Raw      & $-0.499$ & $0.256$ & $5.15\times10^{-2}$  & $7.36\times10^{-2}$  && $1.783$  & $1.839$ & $0.670$              & $0.670$              \\
 & Robust   & $-0.815$ & $0.004$ & $<0.001$             & $<0.001$             && $2.452$  & $0.025$ & $<0.001$             & $<0.001$             \\
 & Trimmed5 & $-0.526$ & $0.264$ & $4.69\times10^{-2}$  & $6.70\times10^{-2}$  && $3.557$  & $2.228$ & $0.251$              & $0.264$              \\
\cmidrule{1-11}
\multirow{3}{*}{White}
   & Raw      & $-0.240$ & $0.254$ & $0.345$              & $0.431$              && $3.789$  & $2.707$ & $0.303$              & $0.319$              \\
 & Robust   & $-0.243$ & $0.003$ & $<0.001$             & $<0.001$             && $5.830$  & $0.038$ & $<0.001$             & $<0.001$             \\
 & Trimmed5 & $-0.284$ & $0.277$ & $0.305$              & $0.381$              && $6.263$  & $3.596$ & $0.143$              & $0.179$              \\
\cmidrule{1-11}
\multirow{3}{*}{Women}
   & Raw      & $-0.149$ & $0.336$ & $0.656$              & $0.691$              && $9.584$  & $1.985$ & $1.53\times10^{-5}$  & $4.36\times10^{-5}$  \\
 & Robust   & $-0.285$ & $0.007$ & $<0.001$             & $<0.001$             && $19.407$ & $0.078$ & $<0.001$             & $<0.001$             \\
 & Trimmed5 & $-0.146$ & $0.346$ & $0.672$              & $0.707$              && $9.439$  & $2.075$ & $4.77\times10^{-5}$  & $1.59\times10^{-4}$  \\
\bottomrule
\end{tabular}
\end{adjustbox}
\end{table}

\begin{table}[H]
\centering
\caption{Full Regression Coefficients --- DeepSeek. Raw logit coefficients, standard errors, and $p$-values for the intercept ($\alpha$) and slope ($\beta$) from logistic regression models of representation, by demographic group and estimation method. $p$-values reported unadjusted and with Benjamini--Hochberg FDR correction.}
\label{tab:coefs_deepseek}
\begin{adjustbox}{max width=\linewidth}
\begin{tabular}{llrrrrrrrrr}
\toprule
& & \multicolumn{4}{c}{\textbf{Intercept ($\alpha$)}} & & \multicolumn{4}{c}{\textbf{Slope ($\beta$)}} \\
\cmidrule{3-6} \cmidrule{8-11}
\multicolumn{1}{c}{\textbf{Group}} &
\multicolumn{1}{c}{\textbf{Method}} &
\multicolumn{1}{c}{\textbf{Coef.}} &
\multicolumn{1}{c}{\textbf{SE}} &
\multicolumn{1}{c}{\textbf{$p$}} &
\multicolumn{1}{c}{\textbf{$p_{\text{FDR}}$}} & &
\multicolumn{1}{c}{\textbf{Coef.}} &
\multicolumn{1}{c}{\textbf{SE}} &
\multicolumn{1}{c}{\textbf{$p$}} &
\multicolumn{1}{c}{\textbf{$p_{\text{FDR}}$}} \\
\midrule
\multirow{3}{*}{Asian}
   & Raw      & $-2.215$ & $0.371$ & $2.37\times10^{-9}$  & $6.77\times10^{-9}$  && $17.745$ & $3.785$ & $9.68\times10^{-6}$  & $3.23\times10^{-5}$  \\
 & Robust   & $-5.043$ & $0.064$ & $<0.001$             & $<0.001$             && $37.431$ & $0.469$ & $<0.001$             & $<0.001$             \\
 & Trimmed5 & $-2.129$ & $0.377$ & $1.58\times10^{-8}$  & $4.52\times10^{-8}$  && $15.557$ & $4.212$ & $5.47\times10^{-4}$  & $1.22\times10^{-3}$  \\
\cmidrule{1-11}
\multirow{3}{*}{Black}
   & Raw      & $-7.714$ & $0.853$ & $1.54\times10^{-19}$ & $7.69\times10^{-19}$ && $26.240$ & $3.741$ & $1.50\times10^{-11}$ & $1.00\times10^{-10}$ \\
 & Robust   & $-10.912$ & $0.800$ & $2.35\times10^{-42}$ & $2.48\times10^{-42}$ && $33.230$ & $3.286$ & $1.03\times10^{-22}$ & $1.03\times10^{-22}$ \\
 & Trimmed5 & $-11.318$ & $1.337$ & $2.57\times10^{-17}$ & $1.45\times10^{-16}$ && $22.813$ & $18.074$ & $0.227$             & $0.253$              \\
\cmidrule{1-11}
\multirow{3}{*}{Hispanic}
   & Raw      & $-0.722$ & $0.294$ & $1.41\times10^{-2}$  & $2.35\times10^{-2}$  && $8.620$  & $2.691$ & $4.63\times10^{-3}$  & $5.79\times10^{-3}$  \\
 & Robust   & $-1.493$ & $0.020$ & $<0.001$             & $<0.001$             && $27.652$ & $0.326$ & $<0.001$             & $<0.001$             \\
 & Trimmed5 & $-0.778$ & $0.318$ & $1.44\times10^{-2}$  & $2.40\times10^{-2}$  && $11.060$ & $3.366$ & $2.80\times10^{-3}$  & $5.60\times10^{-3}$  \\
\cmidrule{1-11}
\multirow{3}{*}{White}
   & Raw      & $0.372$  & $0.263$ & $0.158$              & $0.210$              && $6.849$  & $2.757$ & $3.39\times10^{-2}$  & $3.98\times10^{-2}$  \\
 & Robust   & $0.455$  & $0.012$ & $<0.001$             & $<0.001$             && $10.125$ & $0.132$ & $<0.001$             & $<0.001$             \\
 & Trimmed5 & $0.401$  & $0.278$ & $0.149$              & $0.199$              && $10.952$ & $3.555$ & $5.12\times10^{-3}$  & $8.53\times10^{-3}$  \\
\cmidrule{1-11}
\multirow{3}{*}{Women}
   & Raw      & $-0.169$ & $0.342$ & $0.622$              & $0.691$              && $11.505$ & $2.346$ & $7.51\times10^{-6}$  & $3.00\times10^{-5}$  \\
 & Robust   & $-0.222$ & $0.026$ & $5.60\times10^{-18}$ & $5.60\times10^{-18}$ && $27.606$ & $0.432$ & $<0.001$             & $<0.001$             \\
 & Trimmed5 & $-0.166$ & $0.354$ & $0.639$              & $0.707$              && $11.398$ & $2.452$ & $2.22\times10^{-5}$  & $1.11\times10^{-4}$  \\
\bottomrule
\end{tabular}
\end{adjustbox}
\end{table}

\FloatBarrier
\clearpage
\section{Mixed-Race Representation}
\label{app:mixed_race}

\begin{figure}[htbp]
  \centering
  \includegraphics[width=0.6\textwidth, scale=0.1]{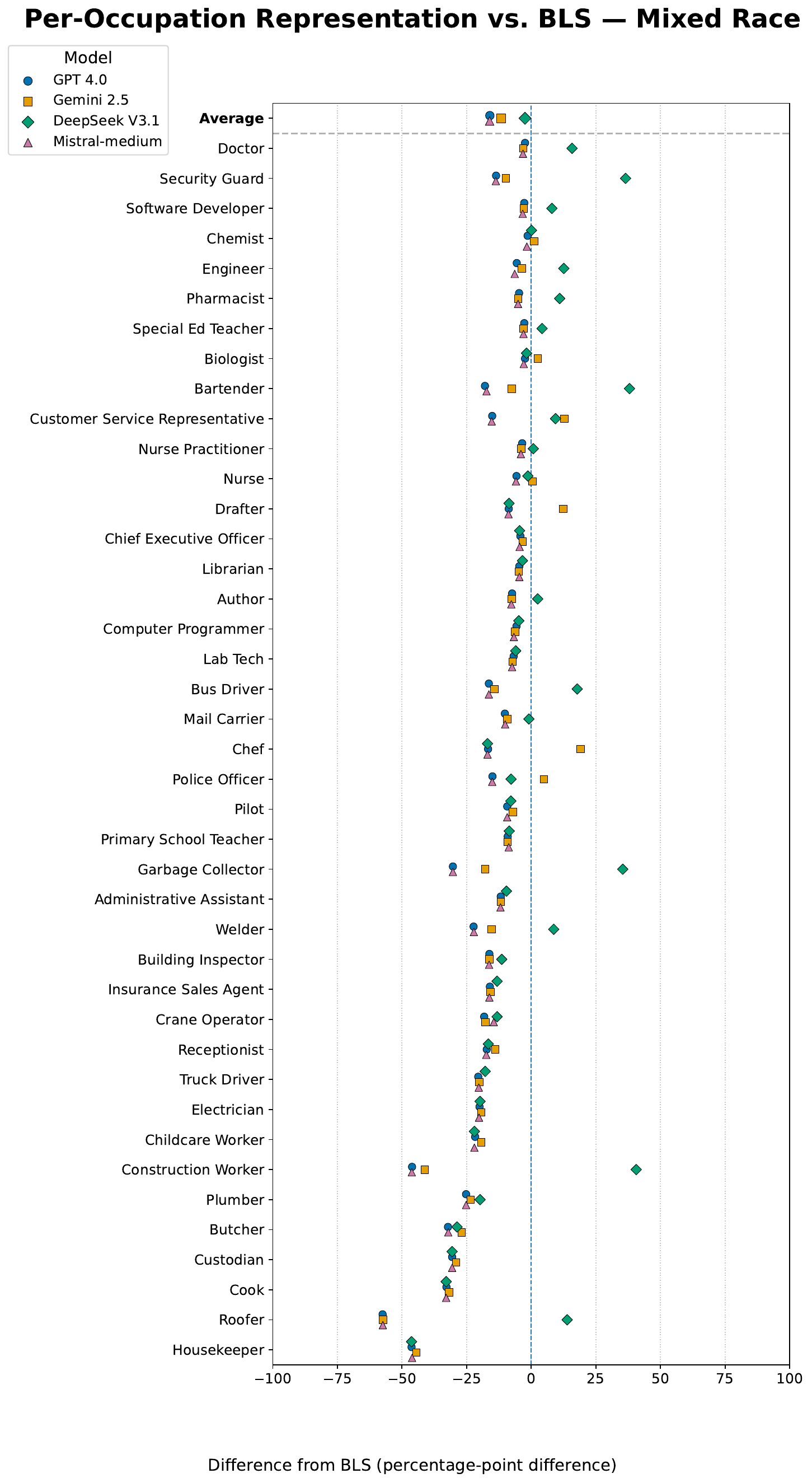}
  \caption{Multi-race representation by occupation and model. Each point shows the percentage-point difference between model-generated multi-race shares and BLS baselines. Across all four models, fewer than 1\% of generated personas receive multi-race labels, resulting in systematic underrepresentation of 10--40 percentage points below BLS. The gap is largest in service and manual occupations where real-world multi-race participation is highest.}
  \label{fig:app_mixed_race}
  \Description{Dot plot showing percentage-point differences in multi-race representation across 41 occupations for four models. Nearly all points fall to the left of zero, indicating that models generate far fewer multi-race personas than exist in BLS data. Gaps range from roughly 10 to 40 percentage points.}
\end{figure}

\paragraph{A note on BLS race and ethnicity definitions.} BLS reports race (White, Black, Asian) and Hispanic ethnicity as overlapping categories rather than mutually exclusive labels: a worker may identify both as Hispanic and as White, Black, or Asian. We operationalize the BLS multi-race baseline as $\max(0,\, p_\text{White} + p_\text{Black} + p_\text{Asian} + p_\text{Hispanic} - 100)$, which captures the implied share of workers reporting more than one of these labels. In practice this share is driven largely by Hispanic-identifying workers who also report a racial category. The term ``multi-race'' should be read in that sense throughout this appendix and the corresponding discussion in the main text. Consequently, the gap reported here reflects models' failure to assign overlapping Hispanic-and-race labels more than their failure to generate intra-race combinations such as Black-and-White, which the underlying BLS occupation tables do not separately report.

\FloatBarrier
\clearpage
\section{Salary Comparisons}
\label{app:salary}
\begin{figure}[!ht]
  \centering
  \includegraphics[width=0.75\textwidth]{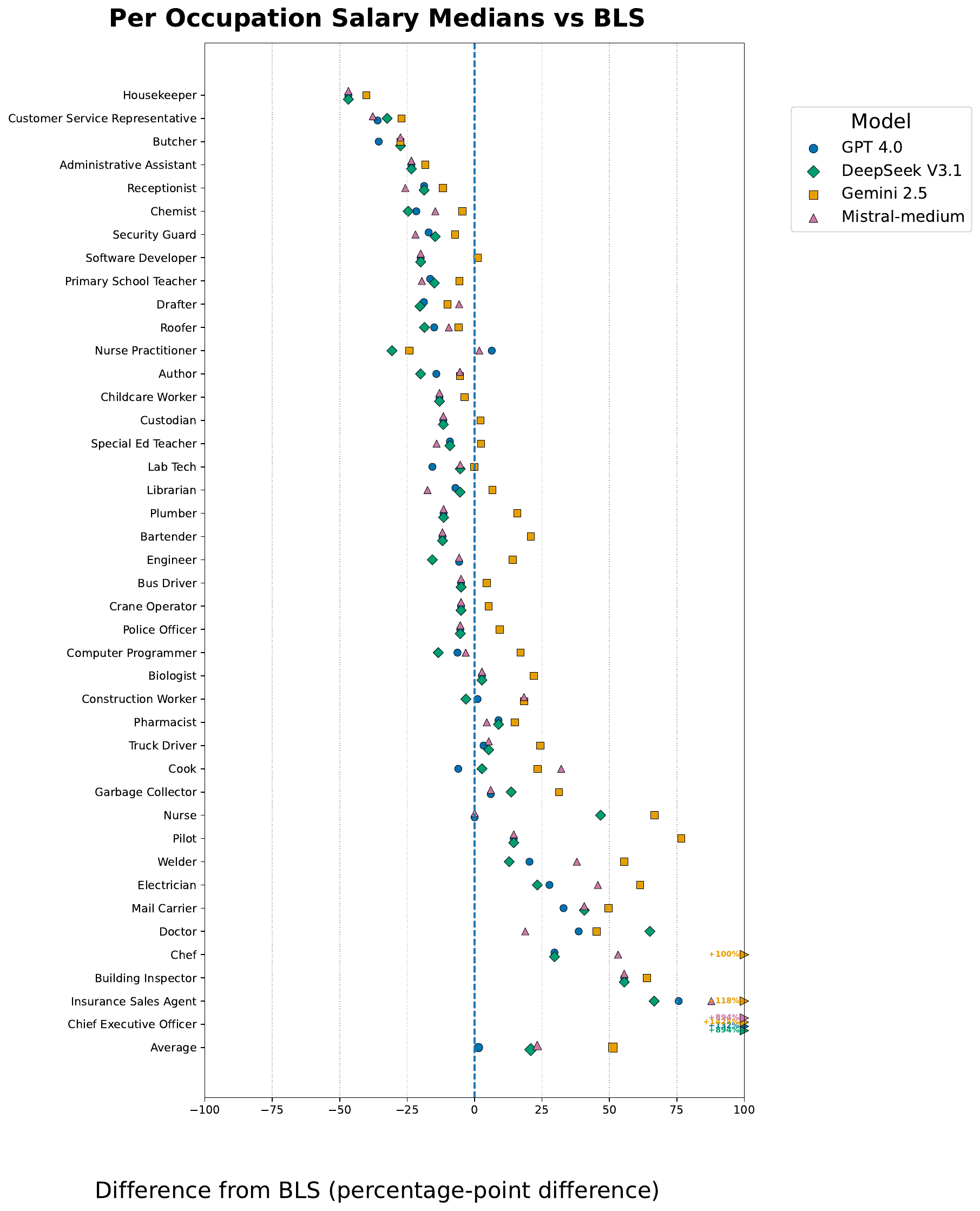}
  \caption{Model-generated median salary compared to BLS median annual earnings by occupation. Each row is an occupation; dots show each model's median generated salary alongside the BLS benchmark. All four models track BLS salary rankings moderately well (Spearman $\rho = 0.73$--$0.80$), though all substantially overestimate salaries for high-prestige occupations such as CEO, where model-generated medians range from +132\% (GPT-4) to +1{,}425\% (Gemini) above BLS; these points exceed the visible x-axis range and are shown as edge arrows with their actual values labeled.}
  \label{fig:app_salary_dotplot}
  \Description{Horizontal dot plot of 41 occupations comparing BLS median salary to model-generated median salaries for four models. GPT-4 dots cluster near BLS values. Other models show wider dispersion, particularly for high-salary occupations where generated salaries exceed BLS by large margins.}
\end{figure}
\begin{figure}[htbp]
  \centering
  \includegraphics[width=0.85\textwidth]{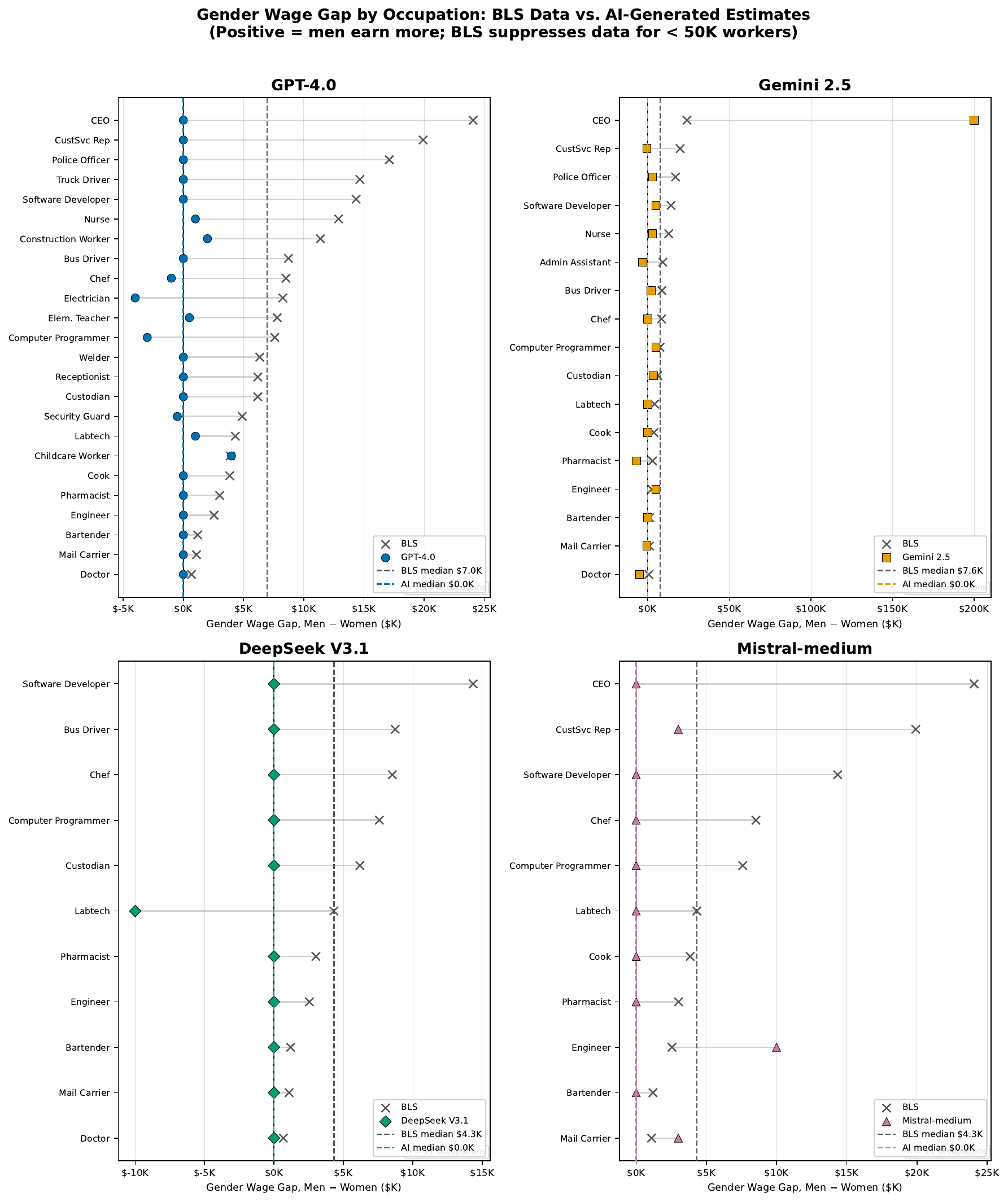}
  \caption{Gender wage gap comparison: BLS versus model-generated salaries. For occupations where both BLS and model data are available by gender, BLS shows a median wage gap of approximately \$7,000 favoring men. All four models compress this gap to a median of \$0, assigning effectively identical salaries to men and women within the same occupation. Analysis is restricted to occupations where BLS reports gender-specific earnings (estimated full-time workforce $\geq$ 50,000) and where the model generates personas of both genders.}
  \label{fig:app_salary_wagegap}
  \Description{Paired comparison showing BLS gender wage gaps and model-generated wage gaps across occupations. BLS bars extend rightward indicating men earn more, while model bars cluster at zero, showing that models erase the real-world gender pay disparity.}
\end{figure}
\end{DIFnomarkup}
\end{document}